\documentclass{aa}  

\usepackage{graphicx}
\usepackage{txfonts}
\usepackage{textcomp}
\usepackage{soul}
\usepackage[normalem]{ulem}
\usepackage{longtable}
\usepackage{hyperref}
\usepackage{xcolor}
\usepackage{caption}

\newcommand{\hii}{H\textsc{ii} region}

\newcommand{\hi}{H\textsc{i}}

\begin{document}

   \title{Deep radio characterization of the supernova remnant G343.1-00.7}

%   \subtitle{I. Overviewing the $\kappa$-mechanism}

   \author{S. Loru
          \inst{1},
           A. Ingallinera\inst{1},
           C. Trigilio\inst{1},
           G. Umana\inst{1},
           D. Uro{\v s}evi{\'c}\inst{2},
           B. Arbutina\inst{2},
           D. Leahy\inst{3},
           C. Bordiu\inst{1,4},
           F. Bufano\inst{1}, 
           I. U. Aalia\inst{5,6},
           C. Buemi\inst{1},
           F. Cavallaro\inst{1},
           E. Egron\inst{7},
           M. D. Filipovi{\'c}\inst{8},
           L. Heino\inst{1},
           A. M. Hopkins\inst{5},
           S. Lazarevi{\'c}\inst{8,9,10}, 
           P. Leto\inst{1},
           A. Pellizzoni\inst{7}, 
           S. F. Rahman\inst{11}, 
           S. Riggi\inst{1},
           A.C. Ruggeri \inst{1} and
           T. Zafar\inst{5,6}
         }

   \institute{$^{1}$ INAF, Osservatorio Astrofisico di Catania, Via Santa Sofia 78, 95123 Catania, IT\\
   \email{sara.loru@inaf.it}\\
   $^{2}$ Department of Astronomy, Faculty of Mathematics, University of Belgrade, Studentski trg 16, 11000 Belgrade, Serbia\\ 
   $^{3}$Department of Physics and Astronomy, University of Calgary, Calgary, AB T2N 1N4, Canada\\
   $^{4}$Instituto de Astrofísica de Andalucía (IAA-CSIC), Glorieta de la Astronomía s/n, E-18008 Granada, Spain\\
   $^{5}$School of Mathematical and Physical Sciences, 12 Wally’s Walk, Macquarie University, NSW 2109, Australia\\
   $^{6}$Astrophysics and Space Technologies Research Centre, Macquarie University, Sydney, NSW 2109, Australia \\   
   $^{7}$INAF, Osservatorio Astronomico di Cagliari, Via della Scienza 5, 09047 Selargius, Italy\\
   $^{8}$Western Sydney University, Locked Bag 1797, Penrith South DC, NSW 2751, Australia\\
   $^{9}$Australia Telescope National Facility, CSIRO, Space and Astronomy, PO Box 76, Epping, NSW 1710, Australia\\ 
   $^{10}$Astronomical Observatory, Volgina 7, 11060 Belgrade, Serbia
   \\
   $^{11}$SBASSE at Lahore University of Management Sciences, LUMS,54792, Lahore, Pakistan\\
}

   \date{}

% \abstract{}{}{}{}{} 
% 5 {} token are mandatory
 
  \abstract
{Supernova remnants (SNRs) are extended radio sources whose morphology and evolution are shaped by the interaction of strong shocks with the surrounding medium. Radio morphological and spectral studies are essential for investigating their local physical properties and particle acceleration mechanisms.
}
{We conducted a detailed radio morphological analysis of the SNR G343.1$-$00.7 and accurately determined its radio continuum spectral properties, both integrated and spatially resolved. 
}
{We combined the 0.944 GHz Australian Square Kilometre Array Pathﬁnder (ASKAP) image from the Evolutionary Map of the Universe (EMU) survey with 0.088–0.200 GHz MWA-GLEAM data to characterize the integrated radio spectrum of G343.1$-$00.7 and perform its first sensitive spatially resolved spectral analysis.
}
{
We derive an integrated spectral index of $\alpha=-0.50\pm0.01$ for G343.1$-$00.7, confirming the previously reported value with significantly improved accuracy.
We redefine the radio morphology of the remnant, providing the first morphological and spectral evidence that the filamentary structures overlapping the nearby \hii\ G343.147$-$00.44 are physically associated with the SNR. Local brightness-brightness analysis further reveals flatter-spectrum filaments along the southeastern shell,
consistent with a possible thermal bremsstrahlung contribution from radiative or partially radiative shocks.
By comparison with \hi\ and CO data, we identify atomic and molecular gas components kinematically consistent with the SNR distance, although no clear morphological evidence of interaction is found. The gas distribution around the \hii, however, suggests the presence of an expanding \hi\ shell partially embedded within its parent molecular cloud. 
Overall, theoretical models based on our radio characterization place G343.1$-$00.7 between the late Sedov phase and the early pressure-driven phase of its evolution, suggesting that different regions may be at different evolutionary stages.

}
{Our results demonstrate the potential of high-resolution ASKAP observations for characterizing the morphology and spatially resolved spectral properties of SNRs, particularly for poorly studied remnants in complex Galactic environments.}

   
   \titlerunning{SNR G343.1-00.7: deep radio study}
   \authorrunning{S. Loru et al.}
 \maketitle

\section{Introduction}
\label{Sec:Introduction}

Supernova remnants (SNRs) originate from the interaction between the stellar ejecta produced by a supernova explosion and the surrounding medium. They are extended and morphologically complex sources observable across the entire electromagnetic spectrum \citep{Dubner_2015,book1}. Their primary observational signature is radio emission, dominated by non-thermal synchrotron radiation produced by relativistic electrons accelerated at the SNR shocks \citep{book2}. This emission typically follows a power-law spectrum, $S_{\nu}\propto \nu^{\alpha}$, with $\alpha \sim -0.5$. However, the study of an increasing number of Galactic SNRs \citep{Green_catalogue_2025} has revealed a dispersion of about $\sim0.2$ around this canonical value, likely related to differences in evolutionary stage, shock acceleration conditions, and interstellar medium (ISM) properties (\citealt{Reynolds_2011}, \citealt{Dubner_2015}, \citealt{Ranasinghe_2023}).

Although synchrotron emission generally dominates the radio spectrum, additional emission mechanisms may contribute significantly, modifying its spectral shape. In particular, a spectral flattening towards high frequencies ($\gtrsim10$~GHz), producing a characteristic `concave-up' spectrum, has been observed in SNRs at different evolutionary stages \citep{Urosevic_2014}. In evolved remnants, particularly composite and mixed-morphology SNRs, this behaviour is commonly interpreted as the result of competing emission mechanisms becoming significant at high frequencies. 
The most widely discussed scenarios include thermal bremsstrahlung, thermal dust emission, and spinning-dust emission associated with dense or molecular environments interacting with the SNR (\citealt{Onic_2012},\citealt{Urosevic_2014}).
By contrast, in young SNRs a concave-up spectrum may arise from non-linear shock modification induced by accelerated particles and is therefore associated with efficient cosmic-ray (CR) production (\citealt{Amato-Blasi2005}, \citealt{Urosevic_2014}, \citealt{Onic_Urosevic_2015}).
Conversely, diffusive shock acceleration (DSA) theory predicts a spectral steepening at high radio frequencies in evolved SNRs due to synchrotron losses suffered by relativistic electrons as they diffuse away from the acceleration regions. The corresponding break frequency provides important constraints on the maximum energy of the accelerated electrons \citep{Loru_2018}. 

The heterogeneity of the physical properties across morphologically complex remnants further complicates this picture, making a spatially resolved characterization through spectral-index mapping essential for identifying variations in particle acceleration conditions and emission processes, as well as evolutionary differences across the remnant.
Spatial variations in the radio spectral index provide valuable diagnostics of the physical conditions across an SNR \citep{2019PASA...36...45H,2019PASA...36...48H}. They can trace changes in particle acceleration efficiency, reveal the presence of distinct emission components or unrelated superimposed sources, and reflect different evolutionary stages within the remnant.
For example, local regions of the blast wave may become radiative earlier where the ambient density is significantly higher than average, while the bulk of the remnant remains in the adiabatic phase \citep{Reynolds_2008}. As a result, both radiative and non-radiative shocks may coexist within the same SNR, potentially giving rise to significant thermal bremsstrahlung emission from radiative regions \citep{Onic_2012,Urosevic_2026}.
A deep and accurate radio characterization therefore provides a valuable 'fingerprint' of SNRs, particularly for poorly studied or newly identified Galactic remnants, placing important constraints on their evolutionary stage, the properties of the surrounding ISM and circumstellar medium, the emission processes occurring in specific regions, and the energetics of the accelerated particles \citep{2022MNRAS.512..265F,2023AJ....166..149F,2024PASA...41..112F,2025PASA...42..104F,2024MNRAS.534.2918S}.

The advent of new-generation radio facilities, both single-dish and interferometric, is providing high-resolution observations of these extended objects, opening the way to detailed spatially resolved spectral studies of a large number of Galactic SNRs (\citealt{Tian_2005}; \citealt{Sun_2011}; \citealt{Loru_2024}; \citealt{Ball_2025}; \citealt{SKABook_2026}). High-resolution spectral-index maps, combined with brightness--brightness analyses of radio images, are crucial for improving our understanding of region-dependent emission mechanisms and for disentangling the different electron populations associated with local shock conditions within the remnant (\citealt{Egron_2017}, \citealt{Castelletti_2021}, \citealt{Loru_2025}, \citealt{2026A&A...710A.308B}).

G343.1-00.7 is a poorly studied Galactic SNR, first identified as such by \citet{Whiteoak_1996} based on 0.843~GHz observations with the Molonglo Observatory Synthesis Telescope (MOST). Although it is listed as a shell-type SNR in the most recent version of the Green catalogue \citep{Green_catalogue_2025}, its morphology appears more complex than a simple shell structure. The MOST image reveals a square-shaped shell with dimensions of approximately $27'\times21'$, with the brightest region located in the southern part of the remnant, composed of three prominent arc-like features.
Adjacent to the northern edge of G343.1$-$0.7 lies another smaller radio bubble ($\sim13'$ in diameter). The presence of mid-infrared emission at 8~$\mu$m and a flat radio spectral index suggests that this adjacent structure is likely of thermal origin \citep{Whiteoak_1996}. A catalogued \hii\ (G343.147$-$00.443, \citealt{Anderson_2014}) overlaps this smaller shell. To date, the MOST image remains the only published radio image of G343.1$-$0.7 available in the literature.

No OH (1720 MHz) maser emission has been detected in association with G343.1$-$0.7 (\citealt{Green_1997}, \citealt{Koralesky_1998}). The distance to the SNR was estimated by \citet{Ranasinghe_2022} to be $4.9 \pm 0.2$~kpc, based on its local standard of rest (LSR) velocity. A comparable value of $4.5$~kpc was also derived by \citet{Pavlovic_2013} using the radio surface brightness–diameter ($\Sigma$–D) relation.
The region encompassing G343.1$-$0.7 has been observed in $\gamma$ rays with both the Fermi-Large Area Telescope (LAT) and the High Energy Stereoscopic System (HESS), but no significant detection was reported \citep{Ferrand_2012}\footnote{\url{http://snrcat.physics.umanitoba.ca}}.

In this work, we present new Australian Square Kilometre Array Pathfinder (ASKAP) observations of G343.1-0.7 at 0.944 GHz, providing the highest-resolution radio image of this remnant published to date. Combined with Murchison Widefield Array (MWA) observations, these data enable us to derive accurate integrated flux-density measurements over 0.155–0.944 GHz, produce the first spatially resolved spectral-index map of the remnant, and perform brightness–brightness analyses of selected regions. This detailed radio characterization allows us to disentangle, for the first time, the emission from the SNR and the adjacent \hii, identify candidate thermal and non-thermal emitting filaments, and place new constraints on the remnant morphology, evolutionary stage, ambient environment, and particle energetics.

In Sect.\ref{sec:Radio data}, we describe the ASKAP and MWA data, while related radio images are presented in Sect.\ref{sec:Results}. 
Section \ref{Sec:The radio spectral index behavior} presents the integrated and spatially resolved spectral analysis, including the morphological interpretation of the remnant and an investigation of its molecular environment. Section \ref{sec:discussion} discusses the evolutionary stage of G343.1-0.7 based on our radio observations and investigates the emission processes operating across the remnant. Section \ref{sec:summary} summarizes our conclusions.

\section{Observations}
\label{sec:Radio data}

\subsection{ASKAP data}
\label{subsec:ASKAP}

We used data acquired with ASKAP as part of the Evolutionary Map of the Universe (EMU; \citealt{Norris_2011}) Phase 2 Pilot survey. Among the observed Galactic fields is the SCORPIO region (\citealt{Umana_2015}; \citealt{Umana_2021}), covering approximately $2 \times 2$~deg$^2$ and centred at Galactic coordinates $l = 343.5^{\circ}$, $b = 0.75^{\circ}$. This region is included in four tiles: \texttt{EMU\_1650-41}, \texttt{EMU\_1718-41}, \texttt{EMU\_1714-46}, and \texttt{EMU\_1644-46}.

All observations within the EMU Phase 2 Pilot were conducted at a central frequency of 0.944~GHz, using the \texttt{closepack36} configuration with a 0.9$^{\circ}$ pitch. Owing to the availability of very short baselines, ASKAP provides images with a maximum theoretical largest angular scale (LAS) of 50~arcmin, ensuring reliable flux density measurements and effective recovery of low-surface-brightness diffuse emission from extended sources with angular scales comparable to this value.

The data products for each individual tile are publicly available via the CASDA data access portal\footnote{\url{https://data.csiro.au/domain/casdaObservation}} under the project code \texttt{AS101}.
The SNR G343.1$-$00.7 lies within tile \texttt{EMU\_1650-41}, which provides a synthesized beam of $\sim 16.7 \times 13.5$~arcsec$^2$. 
We estimate a background RMS noise of $\sim200~\mu$Jy\,beam$^{-1}$ and a standard deviation of $\sim100~\mu$Jy\,beam$^{-1}$ in a region of the tile surrounding G343.1-0.7 and free of contaminating sources. For comparison, the Stokes V standard deviation is $\sim68\mu$Jy\,beam$^{-1}$.

We note that G343.1$-$00.7 was also observed as part of the ASKAP Early Science programme in Bands 1 (central frequency 0.900 GHz), 2 (1.250 GHz), and 3 (1.550 GHz) in 2019 (scheduling blocks SB8838, SB8845, and SB8850). All images were processed using the same restored beam ($14 \times 14$ arcsec$^2$) and aligned on a common pixel grid with a pixel size of 1.5 arcsec. However, these data suffer from inaccurate primary-beam correction because the ASKAP primary beam had not yet been accurately characterized for these observations (\citealt{Ingallinera_2022} and references therein). As a result, the images are affected by significant systematic issues, particularly in the measured flux densities, preventing their use for the analysis presented in this work. These images are presented in Appendix \ref{Appendix}.

\subsection{MWA data}
\label{subsec:MWA}

We employed public maps from the GaLactic and Extragalactic All-sky MWA (GLEAM) survey \citep{Hurley-Walker_2017}, obtained with MWA and accessible via the GLEAM Postage Stamp Service\footnote{\url{http://gleam-vo.icrar.org/gleam_postage/q/form}}, to measure the flux densities of G343.1$-$00.7 in the 0.088–0.200~GHz frequency range. Specifically, we used the four 'wideband' images \citep{Hurley-Walker_2019}, each with a bandwidth of approximately 30~MHz and central frequencies of 0.088, 0.118, 0.155, and 0.200~GHz. These maps offer an angular resolution of $\sim\!100$~arcsec, a sensitivity of 6–10~mJy beam$^{-1}$, and a LAS exceeding $4$~deg.

The GLEAM observations allow us to probe, for the first time, the low-frequency behaviour of the radio spectrum of G343.1$-$00.7. When combined with the EMU data and the three additional flux density measurements available from the literature, they ensure a broad frequency coverage. This significantly improves the accuracy and reliability of the integrated spectral index determination, both for integrated and spatially resolved spectral studies.

\begin{table}
\noindent
	\centering
	\caption{Integrated flux-density measurements of the SNR G343.1$-$00.7.} 
	\label{tab: SNR flux densities}
	\begin{tabular}{|crl|} 
		\hline
		\hline
Freq.	 & Flux density  & Reference    \\
 (GHz)	 & (Jy)\quad\quad\quad &   \\

 \hline
 0.0875	 & $26.6\pm2.2$ & this work \\
 0.1185	 & $23.0\pm1.9$ & this work \\
 0.1545	 & $19.8\pm1.6$ & this work \\
 0.2005	 & $16.8\pm1.3$ & this work \\
 0.843 & $8.5\pm0.6$ & \citet{Whiteoak_1996} \\
 0.944	 & $8.4\pm0.7$ & this work \\
 1.284*    & $5.0\pm0.3$ & \citet{Bordiu_2025} \\
 4.5 & $3.9\pm0.6$ & \citet{Whiteoak_1996} \\
 8.55 & $2.4\pm0.5$ & \citet{Whiteoak_1996} \\

 \hline 
	\end{tabular}
\vspace{1mm}
\begin{minipage}{0.9\linewidth}
\small
\quad* The measurement was not included in the spectral analysis.
\end{minipage}
\end{table}

\begin{table}
\noindent
	\centering
	\caption{Integrated flux-density measurements of the \hii\ G343.147$-$00.44.} 
	\label{tab: HII reg flux densities}
	\begin{tabular}{|crl|} 
		\hline
		\hline
Freq.	 & Flux density  & Reference    \\
 (GHz)	 & (Jy)\quad\quad\quad &   \\

 \hline
 0.0875	 & $2.3\pm0.3$ & this work \\
 0.1185	 & $3.6\pm0.3$ & this work \\
 0.1545	 & $3.1\pm0.4$ & this work \\
 0.2005	 & $3.1\pm0.3$ & this work \\
 0.843 & $4.5\pm0.3$ & \citet{Whiteoak_1996} \\
 0.944	 & $4.4\pm0.4$ & this work \\
 1.284*	 & $3.6\pm0.1$ & \citet{Bordiu_2025} \\
 4.5 & $4.0\pm0.3$ & \citet{Whiteoak_1996} \\
 8.55 & $2.6\pm0.6$ & \citet{Whiteoak_1996} \\

 \hline 
	\end{tabular}
\vspace{1mm}
\begin{minipage}{0.9\linewidth}
\small
\quad* The measurement was not included in the spectral analysis.
\end{minipage}
\end{table}

\section{Results}
\label{sec:Results}
\begin{figure*}
    \centering
    \includegraphics[width=8.2cm]{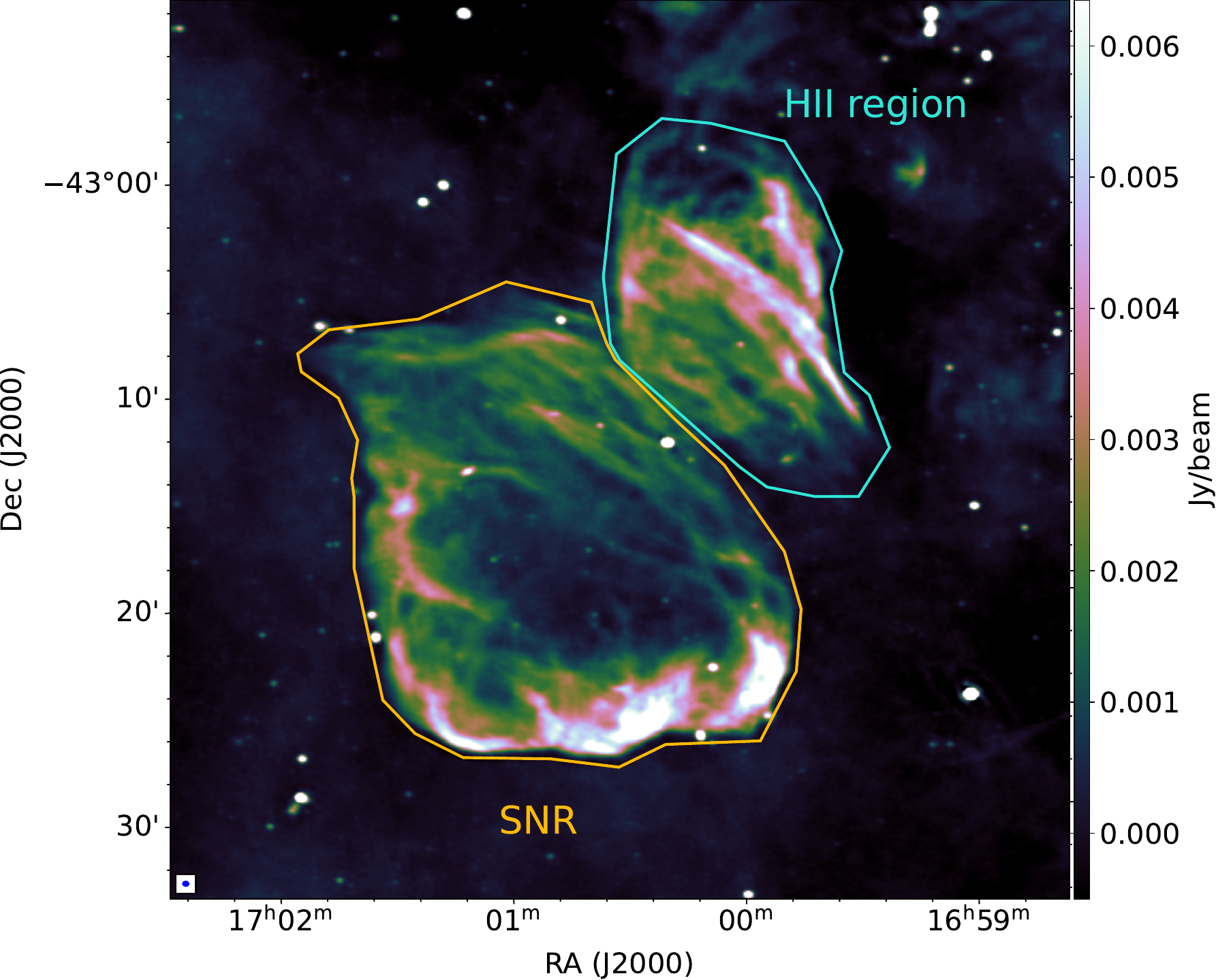}
    \includegraphics[width=8cm]{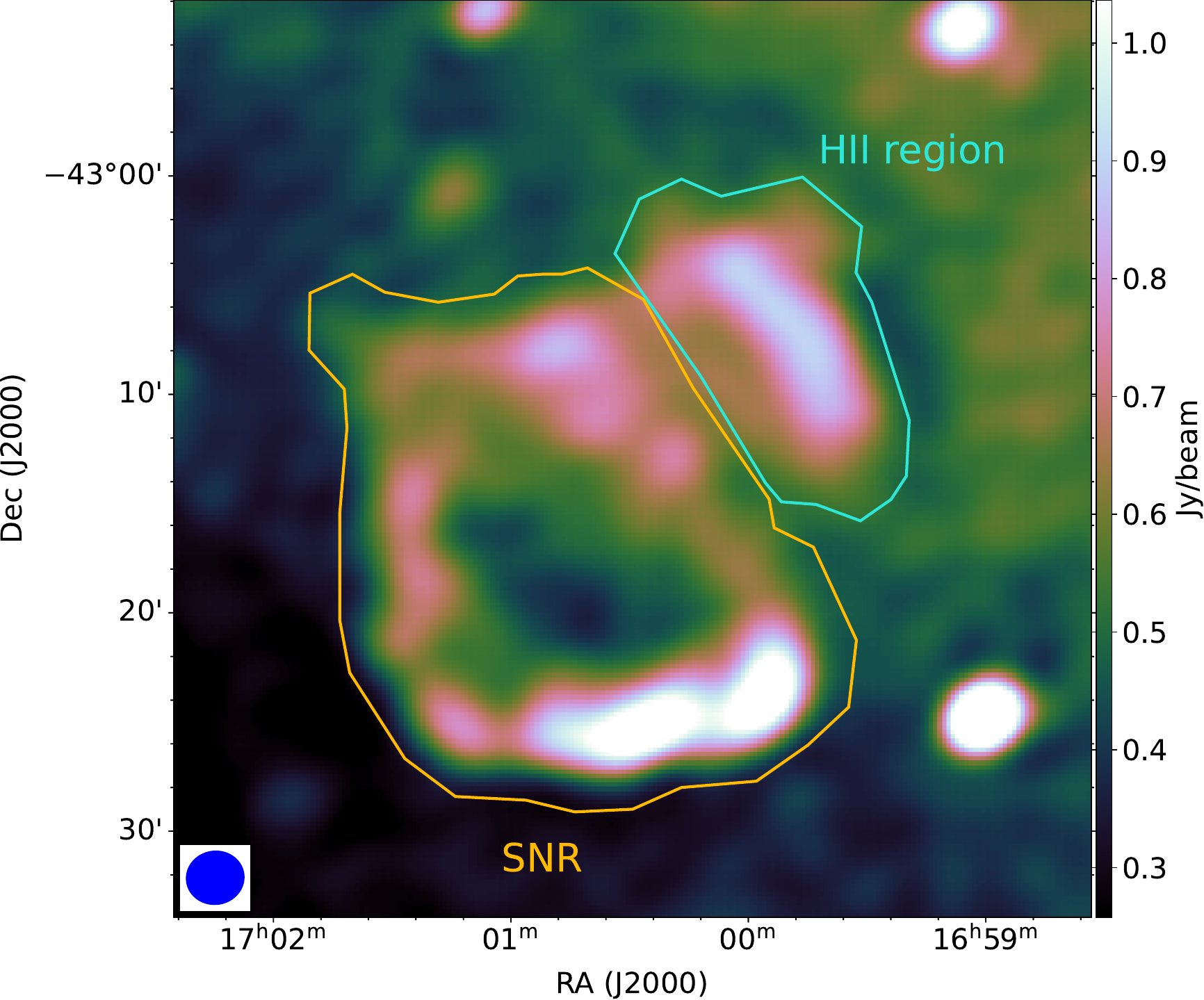}

    \caption{Continuum radio maps of the SNR G343.1$-$00.7 obtained with EMU at $0.944$~GHz (\textit{left})  and GLEAM at $0.200$~GHz (\textit{right}). The beam size is indicated by the blue circle in the bottom-left corner of each map. The orange and cyan outlines in both images indicate the extraction regions used to estimate the flux densities of the SNR G343.1$-$00.7 and the \hii\ G343.147$-$00.443, respectively.}
    
    \label{fig:brightness_maps}
    
\end{figure*}

In Fig.~\ref{fig:brightness_maps}, we present G343.1$-$00.7 and the adjacent \hii\ G343.147$-$00.443 as observed with ASKAP-EMU at 0.944 GHz and MWA-GLEAM at 0.200 GHz. With an angular resolution of $\sim17^{\prime\prime}$, the EMU image provides the highest-resolution view ever published of the G343.1$-$00.7 complex. In this image, the three arc-like structures that define the southern boundary of the remnant are definitely resolved, while two additional, fainter arcs are detected for the first time along the eastern edge of with detection significances above $10\sigma$.

The EMU data also reveal, in unprecedented detail, the filamentary structures extending in a north-east–south-west direction, which make up the more diffuse northern region of the remnant. The \hii\ is also clearly detected, showing a roughly elliptical diffuse emission centred at ($\alpha$, $\delta$) = (17$^{\mathrm{h}}$00$^{\mathrm{m}}$02.7$^{\mathrm{s}}$, $-43^{\circ}04^{\prime}53.7^{\prime\prime}$), with semi-major and semi-minor axes of approximately $7.8^{\prime}$ and $5.4^{\prime}$. It also displays sharper filamentary features aligned north-east–south-west, the brightest of which is centred at ($\alpha$, $\delta$) = (16$^{\mathrm{h}}$59$^{\mathrm{m}}$51.6$^{\mathrm{s}}$, $-43^{\circ}05^{\prime}47.3^{\prime\prime}$). The remarkable correspondence in shape and orientation between the SNR's filamentary structures and those observed in the \hii\ suggests a common origin, offering new insights into the morphology of G343.1$-$00.7.

We measured the flux densities of both sources using aperture photometry on the 0.944~GHz EMU image and the four GLEAM images at 0.088, 0.118, 0.155, and 0.200~GHz. For each object and each map, we defined dedicated extraction regions carefully selected to isolate the emission from the SNR and the \hii, respectively. The regions used to estimate the flux densities of the SNR and the \hii\ in the EMU and GLEAM images are indicated by the orange and cyan regions in Fig.~\ref{fig:brightness_maps}.
In both the EMU and 0.200~GHz GLEAM maps, we observed significant background variation between the areas occupied by the two sources. This is primarily due to the complex environment and the presence of diffuse emission near the Galactic plane. To account for this, we estimated and subtracted the background contribution separately for the SNR and the \hii. Background levels were evaluated by measuring the flux in polygonal regions surrounding each source and scaled according to the ratio of the source area to the background area.

The flux-density uncertainties were computed by adding the calibration and statistical errors in quadrature. The statistical uncertainty was estimated as $\sigma\sqrt{N_{\rm beams}}$, where $\sigma$ is the standard deviation measured over the background region and $N_{\rm beams}$ is the number of beam solid angles contained within the source extraction region. For the calibration uncertainty, we adopted 8\% for the EMU data, while for the GLEAM data we used the values reported by \citet{Hurley-Walker_2017}.
The final flux densities are reported in Tables~\ref{tab: SNR flux densities}~\ref{tab: HII reg flux densities} for G343.1$-$00.7 and the \hii, respectively, along with flux density values available in the literature.

\section{Radio spectral characterization}
\label{Sec:The radio spectral index behavior}

\subsection{Integrated radio spectrum}
\label{subsec:Integrated radio spectrum}

\begin{figure*}
    \centering
    \includegraphics[width=\columnwidth]{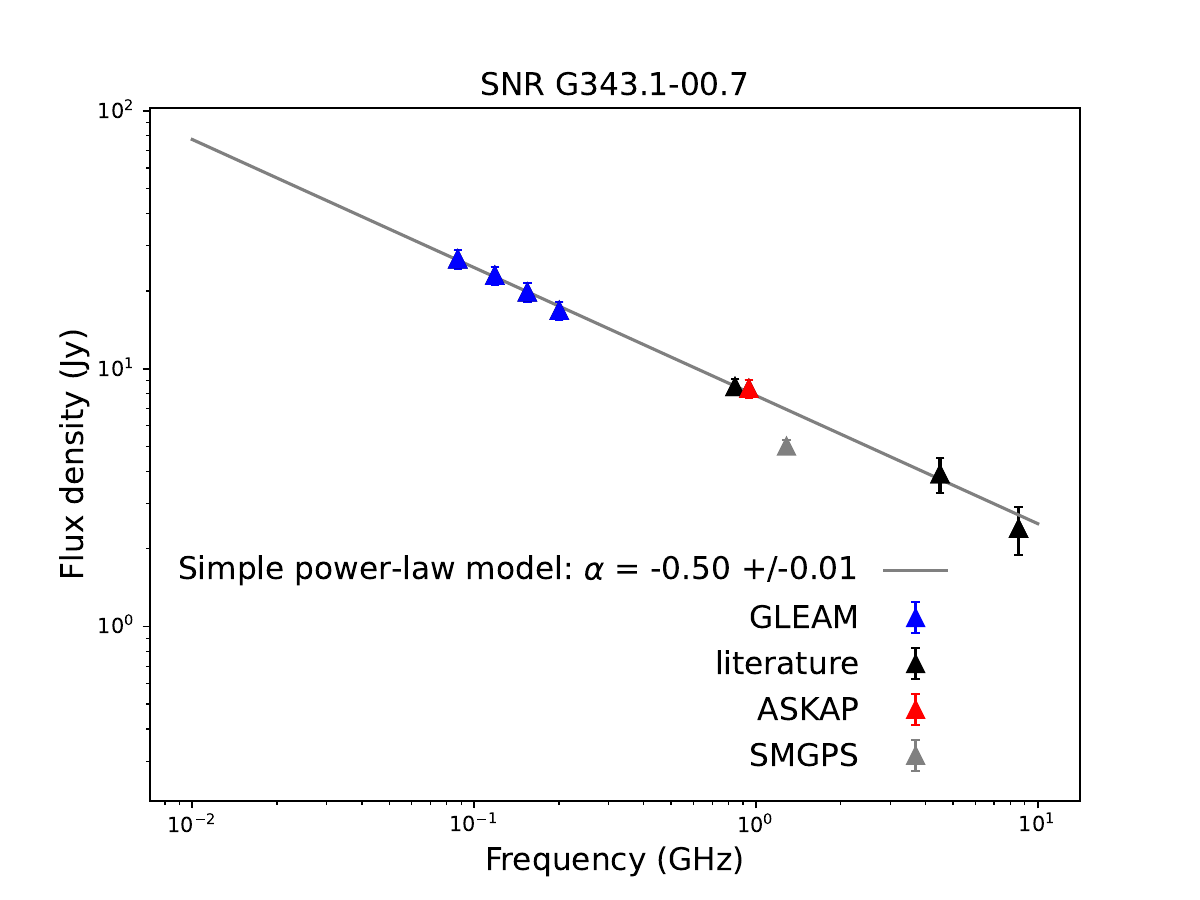}
    \includegraphics[width=\columnwidth]{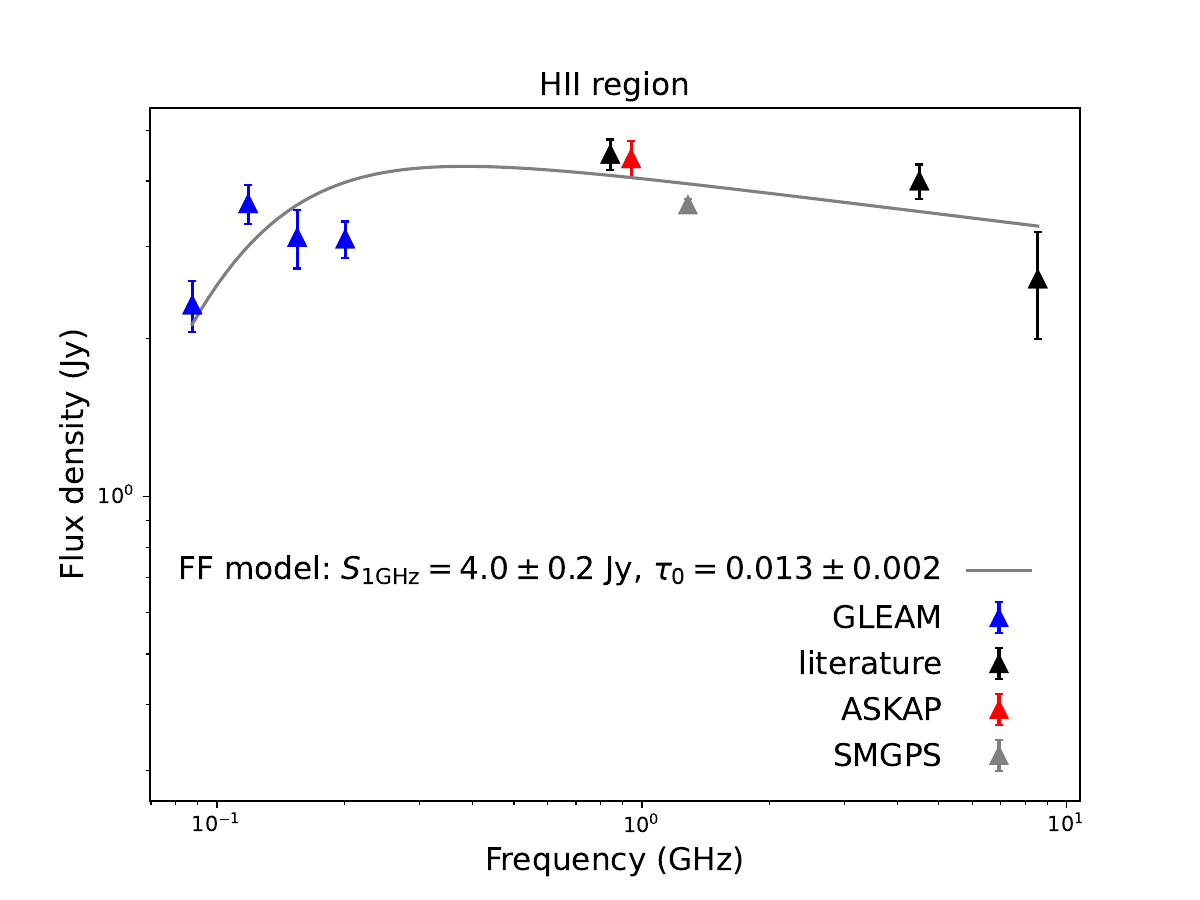}

    \caption{\textit{Left:} SED of SNR G343.1$-$00.7 together with the corresponding weighted least-squares fit. \textit{Right:} SED of the \hii\ G343.147$-$00.443 and the corresponding fit obtained with a simple thermal free--free emission model. In both SEDs (Tables~\ref{tab: SNR flux densities} and \ref{tab: HII reg flux densities}), the SMGPS measurements were excluded from the fitting procedure because of concerns regarding the reliability of their flux density estimates (see Sect.~\ref{sec:Results}).
    }
    \label{fig:SEDs}
    
\end{figure*}

We used all measurements listed in Tables~\ref{tab: SNR flux densities}~\ref{tab: HII reg flux densities} to investigate the integrated spectra of G343.1$-$00.7 and the nearby \hii. For completeness, Tables~\ref{tab: SNR flux densities} and~\ref{tab: HII reg flux densities} also include the MeerKAT SMGPS flux density measurements of G343.1$-$00.7 and the \hii\ G343.147$-$00.443 reported by \citet{Bordiu_2025}. However, the LAS recoverable by the instrument does not ensure reliable flux density recovery for sources with angular sizes larger than $\sim27$~arcmin. This limitation applies to G343.1$-$00.7, which is flagged as 'flux not reliable' in the VizieR catalogue\footnote{G343.1$-$00.7:\url{https://cdsarc.cds.unistra.fr/viz-bin/nph-Cat/html?J/A+A/695/A144/smgpsext.dat.gz}}. Although the SMGPS flux density measurement of the \hii\ is flagged as 'flux reliable' in the corresponding VizieR table\footnote{\hii\ G343.147$-$00.443:\url{https://cdsarc.cds.unistra.fr/viz-bin/nph-Cat/html?J/A+A/695/A144/smgpsext.dat.gz}}, its value is significantly lower than both our measurements and those reported in the literature. Therefore, we excluded the MeerKAT measurements from the spectral analysis. The resulting spectral energy distributions (SEDs) are shown in Fig.~\ref{fig:SEDs}.

For the SNR, we modelled the SED with a simple synchrotron power law ($S_{\nu} \propto \nu^{\alpha}$), yielding an integrated spectral index $\alpha = -0.50 \pm 0.01$, consistent with the previous estimate of $-0.55$ reported by \citet{Whiteoak_1996}. The low-frequency GLEAM data rule out any spectral turnover due to free--free absorption. We also note that our GLEAM and EMU measurements are consistent with the trend defined by previously published data, which we can confirm more robustly thanks to the significantly expanded frequency coverage provided by GLEAM. 

We fitted the \hii\ SED using a simple thermal free-free emission model:
\begin{equation}
S_\nu = S_{1\mathrm{GHz}}\,\nu^2
\frac{1-e^{-\tau_0\nu^{-2.1}}}{1-e^{-\tau_0}},
\end{equation}
where $S_{1\mathrm{GHz}}$ is the flux density at 1~GHz, $\tau_0$ is the optical depth at 1~GHz, and $\nu$ is expressed in GHz. The resulting best-fitting parameters are
$S_{1GHz}=4.0\pm0.2$~Jy and $\tau_0= 0.013 \pm 0.002$. These values imply a transition from the optically thick to the optically thin regime at $\sim0.12$~GHz, consistent with the behaviour expected for a classical, relatively extended \hii.

\subsection{Spatially resolved radio spectrum}
\label{subsec:Integrated radio spectrum}

To investigate possible spectral variations across the two shells, we derived a spectral-index map using the 0.944~GHz EMU and the 0.200~GHz GLEAM images. To account for the different background contributions in the SNR and the \hii, we generated two distinct spectral-index maps starting from their respective background-subtracted images.   
We used the \textsc{CASA} tasks \textsc{convolve2d} and \textsc{imregrid} to convolve and regrid the 0.944~GHz EMU map so that it matched the beam size, coordinate system, and pixel scale of the 0.200~GHz GLEAM image. 

Spectral-index uncertainties were computed through standard error propagation, and we masked the final spectral-index maps where the uncertainty exceeded 0.2 to ensure reliable values.
To preserve an overall view of the complex SNR+\hii\ system, we combined the two maps into a single image, where the pixels in the narrow region at the interface of the two sources were computed as the mean of the corresponding overlapping pixel values. 

\begin{figure}
    \centering
    \includegraphics[width=\columnwidth]{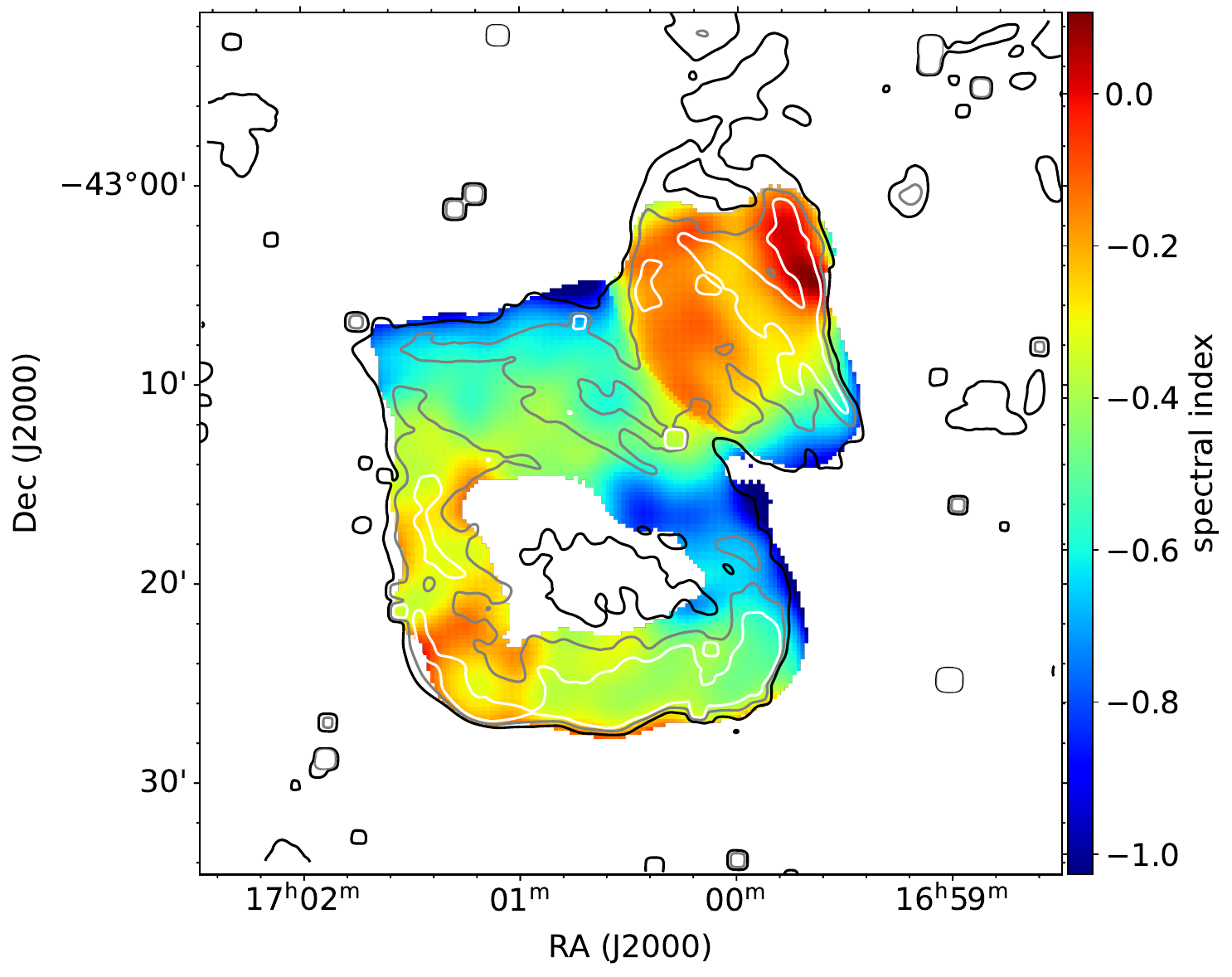}
    \includegraphics[width=\columnwidth]{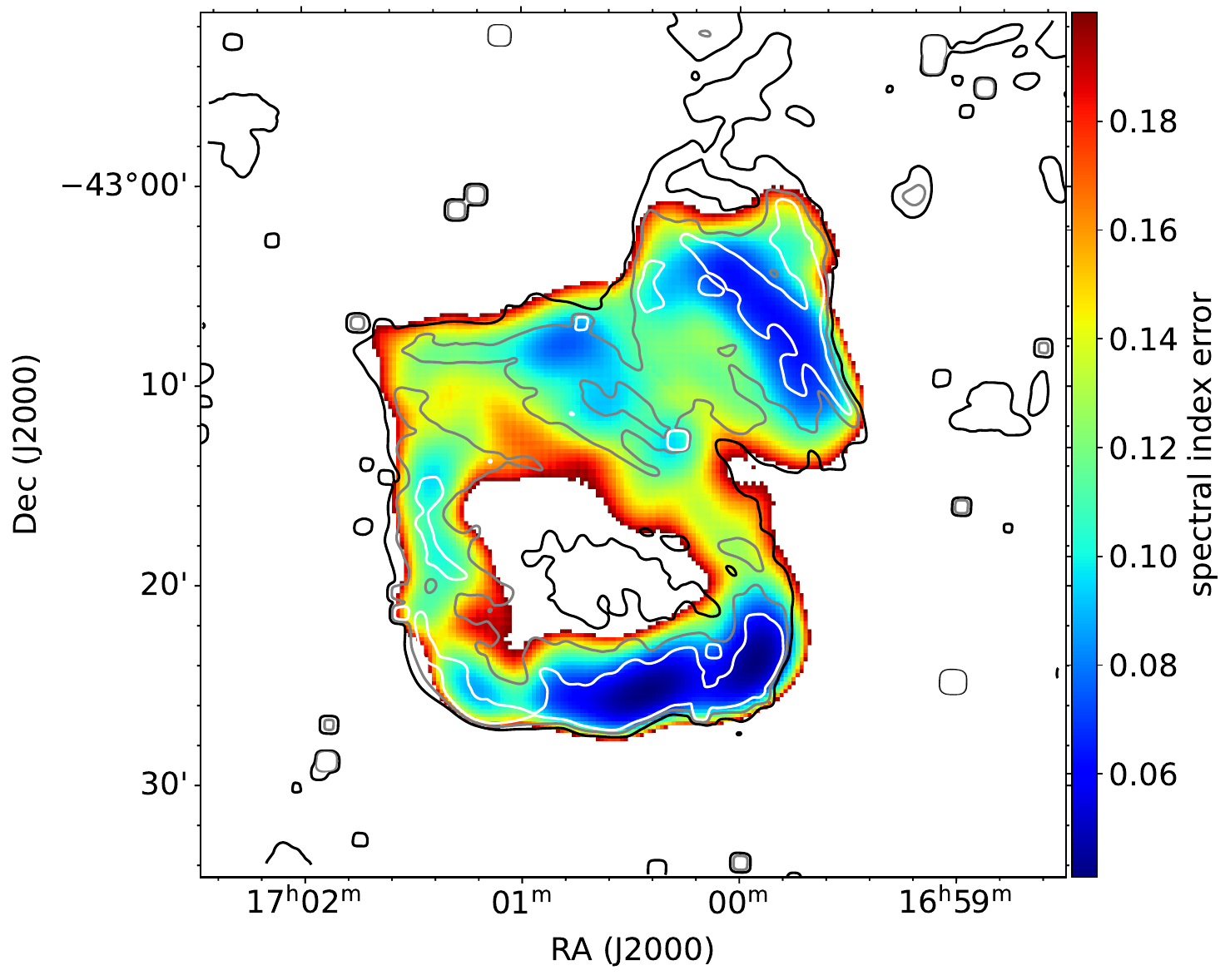}    
    \caption{Spectral-index map of G343.1$-$00.7 (\textit{top}) and related error map (\textit{bottom}) obtained from the GLEAM map at 0.200 GHz and EMU map at 0.944 GHz. The contours indicate the intensity levels of the EMU map at 0.0004, 0.0015, and 0.003~Jy beam$^{-1}$.}
    \label{fig:spix}
\end{figure}

The resulting spectral-index map, together with the associated error map, is shown in Fig.~\ref{fig:spix}. It clearly distinguishes the overall spectral behaviour of the non-thermal emission from the SNR and the thermal emission from the \hii. Significant spectral variations are also detected across both sources.
In particular, the more diffuse and fainter northwestern region of G343.1$-$00.7 exhibits a significantly steeper spectrum, with a mean spectral index of approximately $\alpha \sim -0.63$, compared to the brighter southwestern shell, which shows a mean value of $\alpha \sim -0.34$. Within this region, the two brightest arc-like filaments also appear to have a slightly steeper spectrum (mean $\alpha \sim -0.40$) than the remainder of the region. 
We also point out that the southeastern part of the shell shows flatter spectral indices than the rest of the main shell. Other minor spectral features, such as knots, are also visible in this region, although they do not show a clear correlation with the brightness distribution. 

Significant spectral variations are also observed across the \hii. While much of the north-northeastern part shows spectral index values between $-0.1$ and $0.01$, as expected for thermal emission, the region associated with the brightest filament exhibits a noticeably steeper spectrum (with $\alpha \sim -0.26$). Furthermore, an area with even lower values ($\alpha \sim -0.64$) extends towards the southern part of the region. This spectral behaviour suggests that, in this area, the \hii\ emission coexists with non-thermal features, most likely associated with the nearby SNR G343.1$-$00.7. Using a simple two-component model consisting of optically thin free-free emission ($\alpha_{\rm th}=-0.1$) and synchrotron emission ($\alpha_{\rm nt}=-0.5$), we find that the observed map-based spectral index of the bright filament ($\alpha \simeq -0.26$) is reproduced by a non-thermal contribution of approximately 32\% of the total flux density at 1 GHz, increasing to about 48\% at 0.200 GHz. 

\begin{figure*}
    \sidecaption
    \includegraphics[width=12cm]{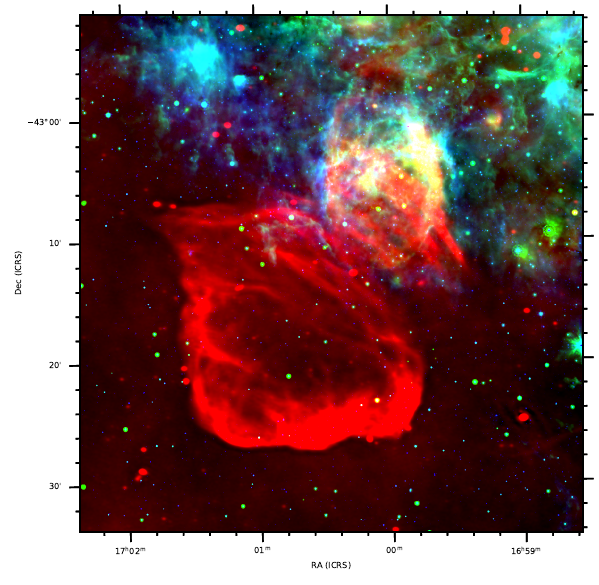} 
    \caption{Three-colour image of G343.1$-$00.7 and the \hii\ G343.147$-$00.443. Red: ASKAP at 0.944 GHz. Green: MIPSGAL at $24$~$\mu$m (resolution of $\sim6^{\prime\prime}$) . Blue: GLIMPSE at $8$~$\mu$m (resolution of $\sim3^{\prime\prime}$).
    }
    \label{fig: rgb}
    
\end{figure*}

\begin{figure*}
    \sidecaption
    \vbox{
        \hsize=12cm
        \centering
  \includegraphics[width=12cm]{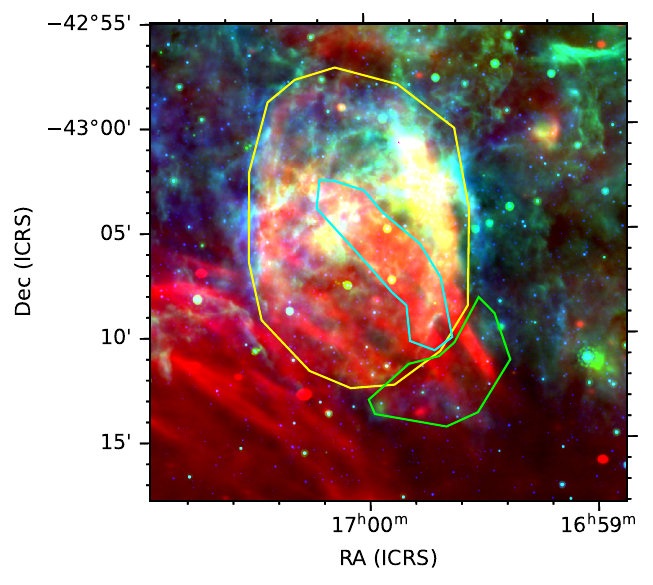}
        \vskip 0.15cm
        \hbox to 12cm{
    \includegraphics[width=5.7cm]{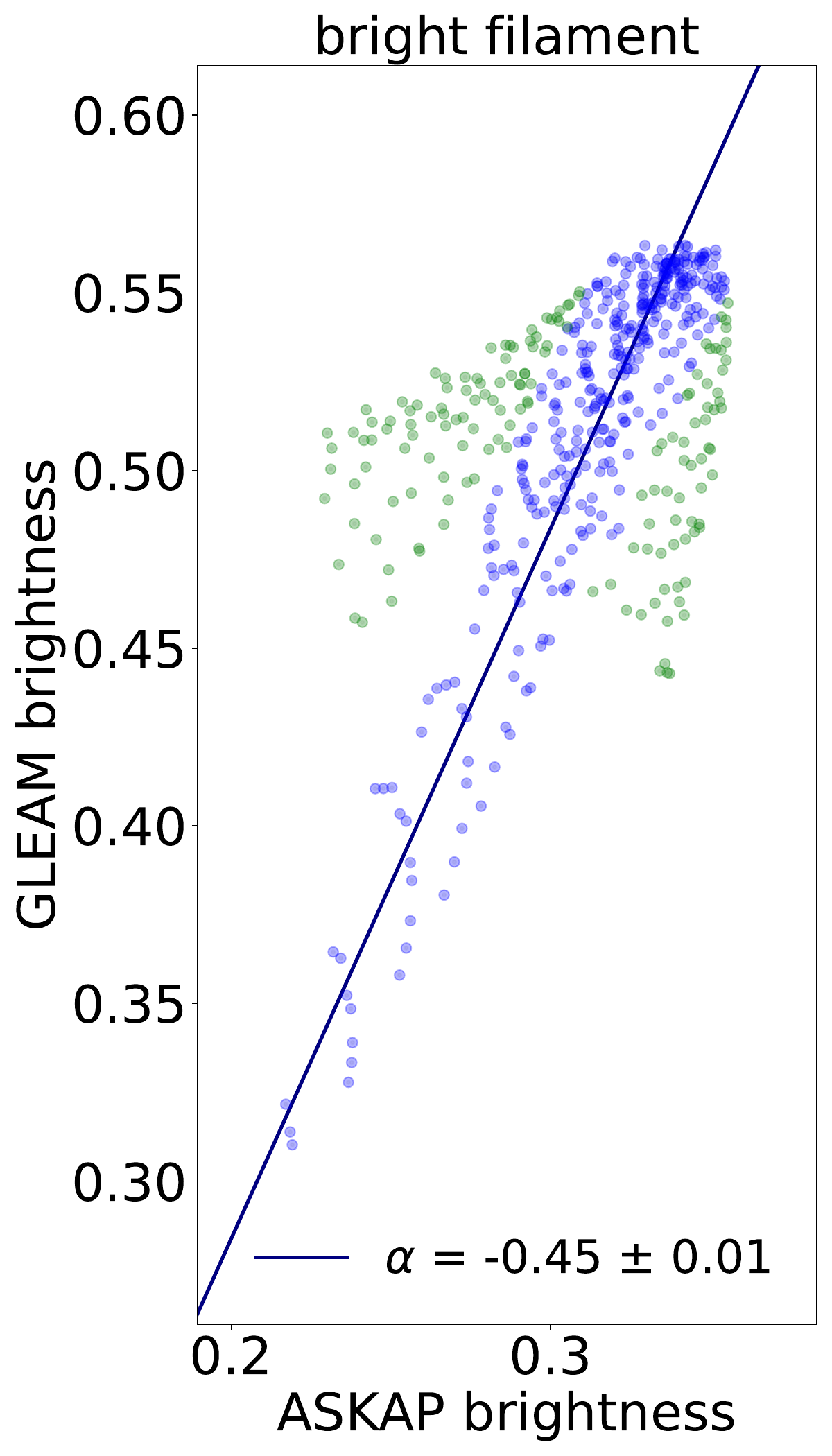}
            \hfill
 \includegraphics[width=5.7cm]{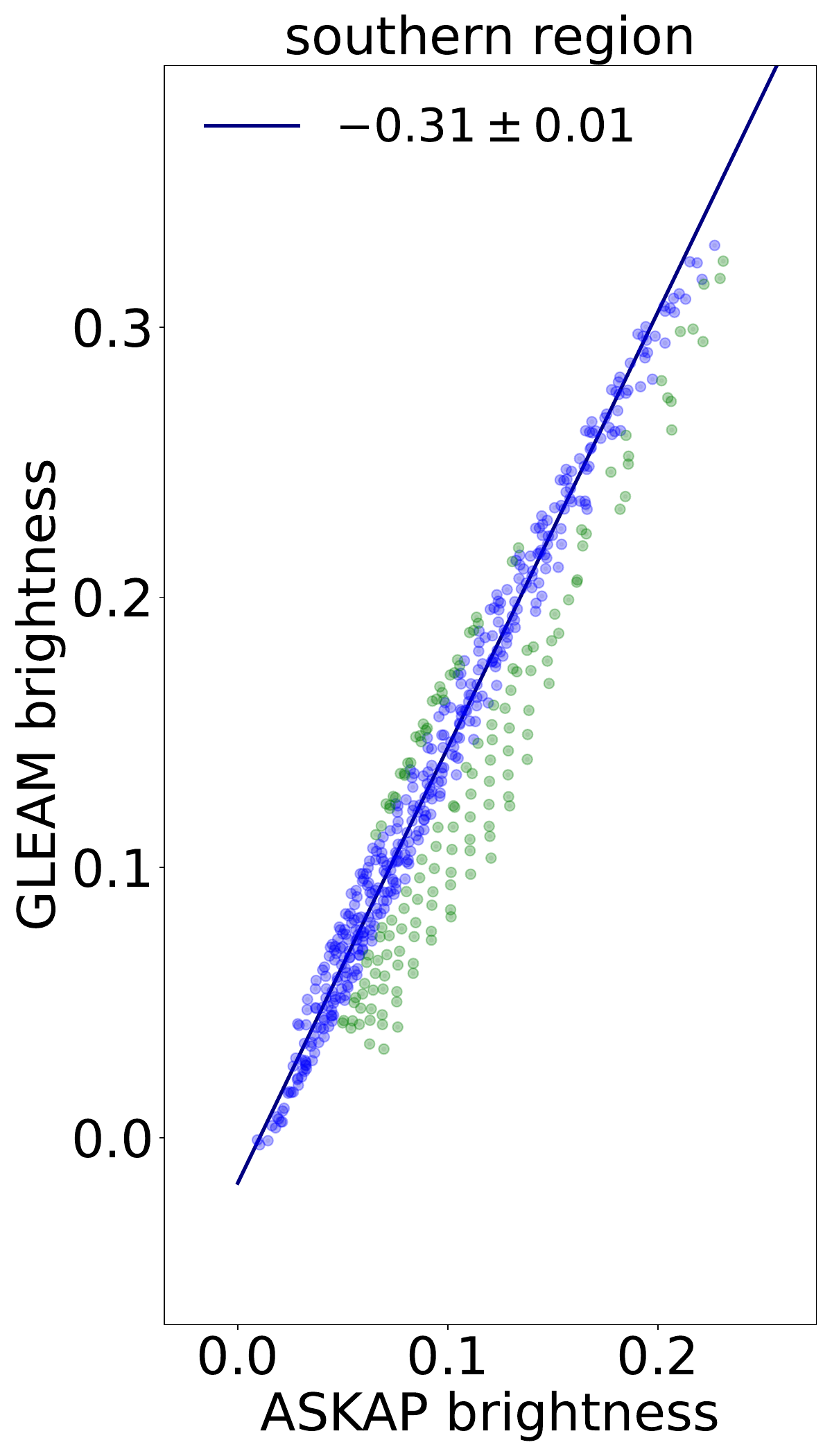}
        }
    }
    \caption{Zoom-in of the three-colour image presented in Fig.~\ref{fig: rgb}, 
    focusing on the northern shell (\textit{upper panel}). The yellow region 
    outlines the \hii, as identified from the radio--infrared morphology, while 
    the cyan and green regions correspond to the bright filament and the southern 
    part of the shell, respectively. Brightness--brightness plots between the EMU 
    and GLEAM data are shown for the bright filament (\textit{bottom left}) and 
    the southern part of the \hii\ (\textit{bottom right}). The blue points 
    represent the inliers identified by the RANSAC algorithm and used for the 
    linear fit, while the green points correspond to the outliers.}
    \label{fig:bb-plots}
\end{figure*}

\begin{figure}
    \centering

    \includegraphics[width=\columnwidth]{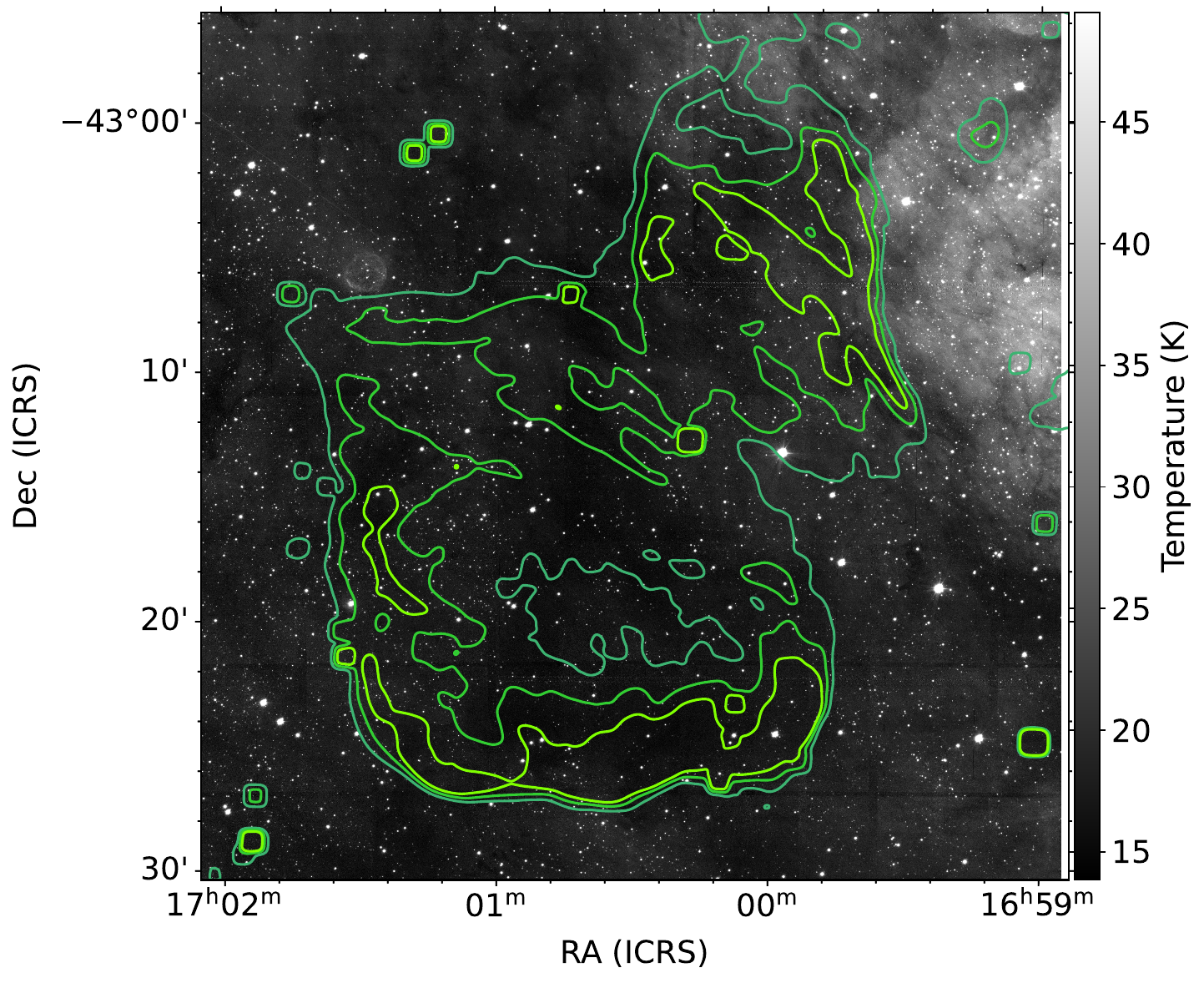}

    \caption{VPHAS+ $H\alpha$ greyscale map towards G343.1$-$00.7. The contours trace the 0.944~GHz radio emission observed with ASKAP at levels of 0.0004, 0.0015, and 0.003~Jy~beam$^{-1}$. No $H\alpha$ filaments spatially coincident with the G343.1$-$00.7 radio emission are detected.}
    
    \label{fig:Halpha}
    
\end{figure}

\subsection{Discerning the G343.1$-$00.7 complex morphology}
\label{Sec:5}

To discern the nature of the emission from the northern shell of the G343.1$-$00.7 complex and to correctly define the morphology of the SNR and the \hii, we exploited the high resolution of the EMU image to perform a detailed radio–IR morphological comparison. We produced a three-colour image showing the radio emission at $0.944$~GHz in red, the 24~$\mu$m mid-IR emission from the Multiband Imaging Photometer for Spitzer Galactic plane survey (MIPSGAL; \citealt{Carey_2009}) data in green, and the 8~$\mu$m IR emission from the Galactic Legacy Infrared Mid-Plane Survey Extraordinaire (GLIMPSE; \citealt{Churchwell_2009}) data in blue. In the resulting image shown in Fig.~\ref{fig: rgb}, the morphology of the \hii\ along the northern shell is clearly outlined by an ellipse centred at ($\alpha$, $\delta$) = (17$^{\mathrm{h}}$00$^{\mathrm{m}}$02.65$^{\mathrm{s}}$, $-43^{\circ}04^{\prime}53.65^{\prime\prime}$), with semi-major and semi-minor axes of approximately $7.8^{\prime}$ and $5.4^{\prime}$, respectively. Here the typical radio–infrared correspondence is met: the 24~$\mu$m and radio emissions are spatially coincident, while the 8~$\mu$m emission appears as a surrounding envelope \citep{Ingallinera_2014}. 
The radio-IR comparison also highlights the morphological mismatch between the \hii\ and the bright radio filaments in the northern shell, which instead retain a remarkably consistent morphology with that of the SNR.

To further investigate the nature of these filaments, we performed a spectral analysis by producing 'BB-plots', i.e. pixel-by-pixel scatter plots between the EMU and GLEAM brightness maps, for different macro-regions of the northern shell.
Fitting the BB-plot points allows us to identify potential spectral features in regions where distinct emission components, associated with different physical processes, overlap. This approach proves particularly effective in cases where such processes coexist within the same region, as it provides a more reliable estimate of the spectral index than that obtained by directly averaging the spectral-index map. In fact, the latter may fail to disentangle the multiple spectral features that are often spatially blended.
As shown by \citet{Loru_2024}, BB-plot analysis represents a crucial tool for disentangling and correctly estimating the spectral indices of SNRs and nearby or co-spatial sources, such as \hii\ or pulsar wind nebulae. 

We produced BB-plots for the brightest radio filament and the southwestern edge of the minor shell, as indicated in Fig.~\ref{fig:bb-plots} (upper panel) by the cyan and green contours, respectively. 
We fitted them using the RANSAC Regressor algorithm \citep{Fischler_1981}, as implemented in the Python library \textsc{scikit-learn}\footnote{\href{https://scikit-learn.org/stable/}{https://scikit-learn.org/stable/}} \citep{scikit-learn}. This algorithm iteratively identifies inliers and outliers to robustly fit a linear model, making it particularly suitable for data affected by noise or contamination. It also allows the identification of possible distinct linear trends within the overall data distribution. 
In Fig.~\ref{fig:bb-plots} (bottom panels), we present the BB-plots and the corresponding linear fits, from which we derive spectral indices of $\alpha = -0.45 \pm 0.01$ for the filament and $\alpha = -0.31 \pm 0.01$ for the southwestern edge of the shell. These values unequivocally confirm the non-thermal nature of both structures. 
We point out that in both BB-plots, the fit of the scatter points shows a significant offset from the origin. This is an expected effect, as both structures are superimposed on the more diffuse emission from the \hii, which acts as a background component in the BB-plot analysis. The spectral indices derived from the BB-plot analysis are consistent with the interpretation presented at the end of Sect.~\ref{subsec:Integrated radio spectrum}, based on the non-thermal fractions inferred from the two-component mixing model applied to the spectral-index map. Together, these results support a scenario in which the bright filament and the southwestern edge are intrinsically dominated by synchrotron emission, while the flatter spectral index measured in the spectral-index map arises from its superposition with the diffuse thermal emission of the H II region.

To further characterise the morphology of G343.1$-$00.7, we also searched for Balmer-dominated filaments tracing the shock front by inspecting public VST Photometric H$\alpha$ Survey of the Southern Galactic Plane and Bulge (VPHAS+; \citealt{Drew_2014}) data in the H$\alpha$ filter. In Fig.~\ref{fig:Halpha}, we compare the H$\alpha$ emission map of the G343.1$-$00.7 region with the EMU radio image. Given the location of the G343.1$-$00.7 complex very close to the Galactic plane, this non-detection is most likely due to strong H$\alpha$ extinction and therefore does not provide meaningful constraints on the nature of the shock.

We searched for a possible X-ray counterpart of G343.1$-$00.7 using the public eROSITA-DE data from Data Release~1 (DR1)\footnote{\url{https://erosita.mpe.mpg.de/dr1/erodat/}}. We inspected the tile numbered~139108, which includes the position of G343.1$-$00.7, and produced a three-colour image combining the 0.2–0.6~keV, 0.6–2.3~keV, and 2.3–5.0~keV energy bands. No X-ray excess emission associated with the SNR was detected.

\subsection{H\textsc{i} and CO environment}
\label{HI environment}

\begin{figure*}
    \centering
    \includegraphics[width=18cm]{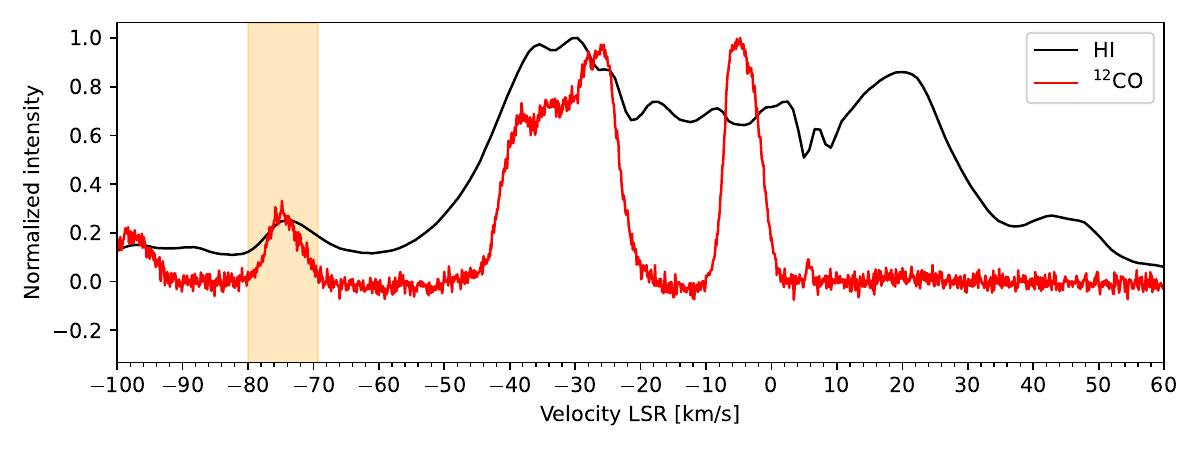}
    
    \includegraphics[width=18cm]{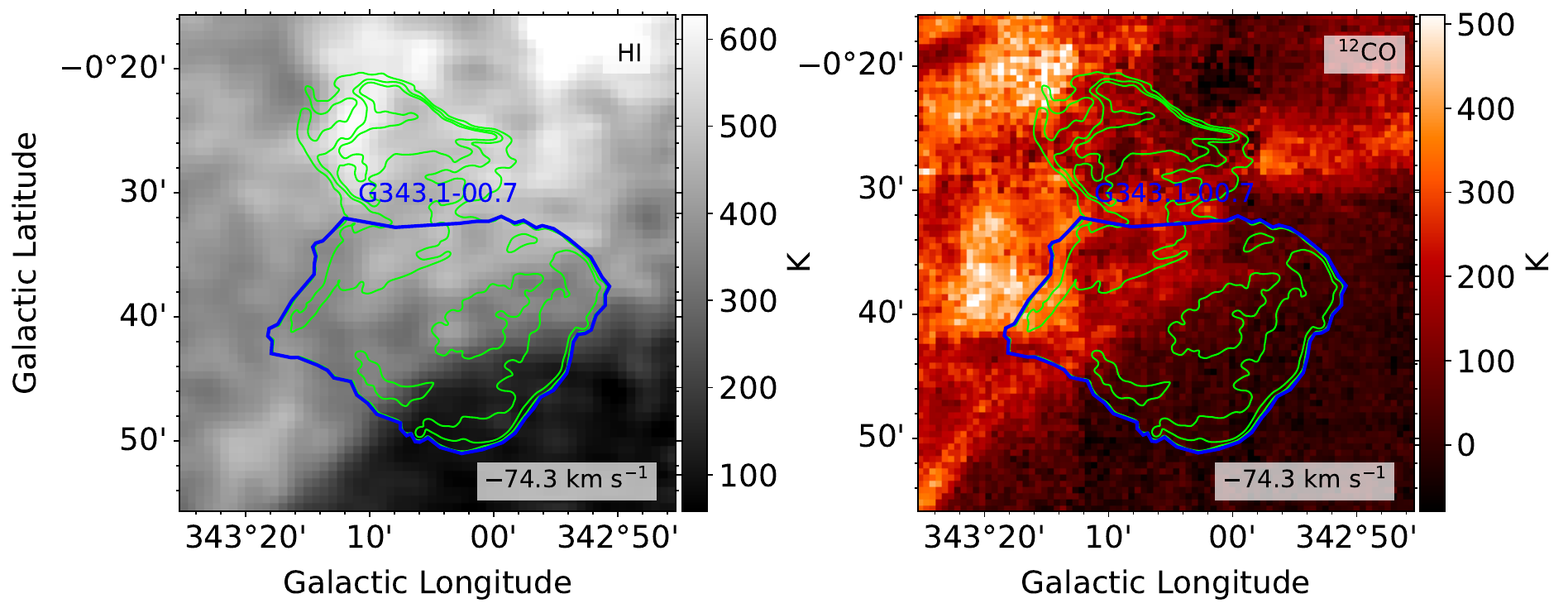}
    
    \caption{\textit{Top:} H\textsc{i} velocity profile extracted from the SNR G343.1$-$00.7 (blue polygon in the bottom panels). The shaded area marks the velocity range from $-80.0$ to $-68.60$~km s$^{-1}$, corresponding to H\textsc{i} gas and $^{12}$~CO molecular material at the same kinematic distance of the SNR. \textit{Bottom:}
    H\textsc{i} (\textit{left}) and $^{12}$~CO (\textit{right}) intensity maps of the G343.1$-$00.7 complex from SGPS and MOPRA data, respectively, integrated over the ($-80.0$,$-68.60$)~km s$^{-1}$ interval.
    ASKAP radio contours at 0.0004, 0.0015, and 0.003~Jy beam$^{-1}$ are shown in green. The blue polygon encloses the SNR as observed in our radio images.}
    \label{fig:vel_prfile_HI_SNR}    
\end{figure*}

\begin{figure*}
    \centering
    \includegraphics[width=18cm]{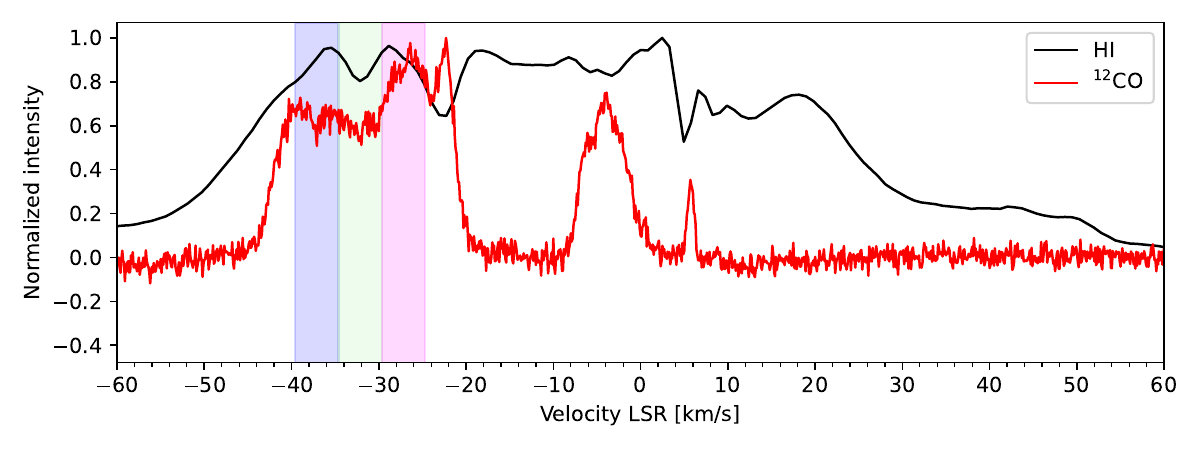}
    
    \includegraphics[width=18cm]{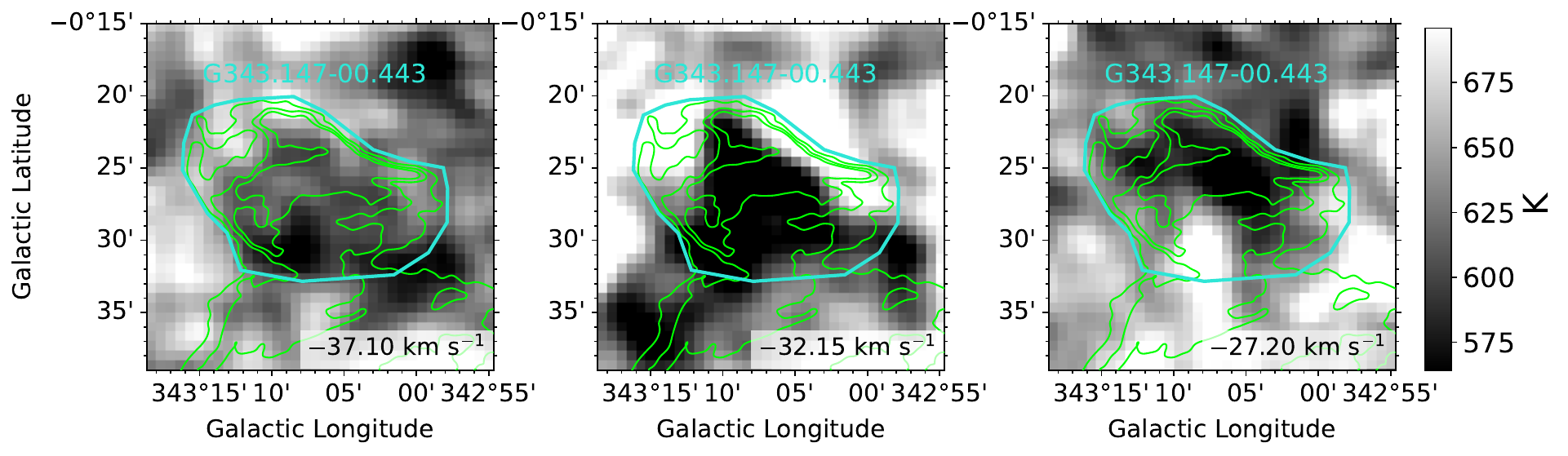}
    \includegraphics[width=18cm]{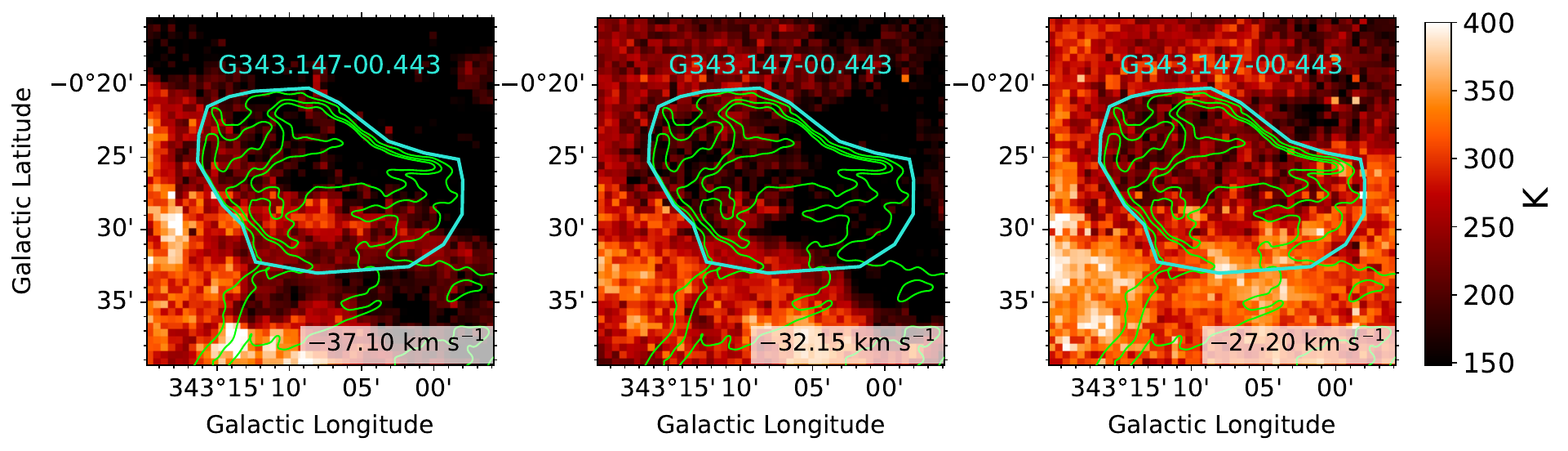}
    \caption{\textit{Top:} H\textsc{i} and $^{12}$~CO velocity profiles extracted from  \hii\ G343.147$-$00.443 (cyan polygon). The shaded area marks the velocity range from $-39.57$ to $-24.7$~km s$^{-1}$, corresponding to H\textsc{i} and $^{12}$~CO-emitting material at a kinematic distance consistent with that of the \hii. 
   The H\textsc{i} (\textit{centre}) and $^{12}$~CO (\textit{bottom}) intensity maps of the \hii\ G343.147$-$00.443 from SGPS and MOPRA spectral-line data, respectively, are integrated over the three velocity ranges highlighted in the top panel. \textit{Left:} $-39.57$ to $-34.63$~km s$^{-1}$ (light-blue). \textit{Centre:} $-34.63$ to $-29.68$~km s$^{-1}$ (light-green). \textit{Right:} $-29.68$ to $-24.73$~km s$^{-1}$ (light-magenta). ASKAP radio contours at 0.0004, 0.0015, and 0.003~Jy beam$^{-1}$ are shown in green. The cyan polygon encloses the \hii\ as identified from the radio images.}
    \label{fig:vel_prfile_HI_HIIreg}    
\end{figure*}

We investigated the molecular environment around the G343.1$-$00.7 complex to assess its possible association with the morphological and spectral properties of both the SNR and the \hii. For this purpose, we compared the ASKAP-EMU radio map with the $^{12}$CO ($J=1$–0) emission from the public MOPRA dataset \citep{MOPRA_2023} and the H\textsc{i} emission from the public SGPS (Southern Galactic Plane Survey, \citealt{SGPS_2005}) spectral-line data.

\subsubsection{SNR G343.1$-$00.7} 
\label{subsubsection:SNR G343.1-00.7}

Starting from the MOPRA and SGPS data cubes, we analysed the $^{12}$CO and H\textsc{i} velocity profiles extracted over the region encompassing the remnant, searching for velocity components morphologically and kinematically associated with G343.1$-$00.7.
Both spectra exhibit an emission peak centred at $\sim -74.3$~km\,s$^{-1}$ (orange shaded area in Fig.\ref{fig:vel_prfile_HI_SNR}, upper panel), corresponding to a near kinematic distance of $5.09^{+0.54}_{-0.25}$~kpc, derived using the Galactic rotation model of \citet{Reid_2014}\footnote{Kinematic distance calculated using the \href{https://www.treywenger.com/kd/}{Kinematic
Distance Calculation Tool}, based on \citet{2018Wenger}.}. This value is consistent with the remnant’s distance of $4.9 \pm 0.2$~kpc derived by \citet{Ranasinghe_2022}.
 
In Fig.~\ref{fig:vel_prfile_HI_SNR}, we present the integrated intensity map of the H\textsc{i} (bottom-left) and $^{12}$CO (bottom-right) emission towards G343.1$-$00.7 in the velocity range $-80.0$ to $-68.60$~km s$^{-1}$, overlaid with radio intensity contours from the ASKAP image at 0.944~GHz. 
The H\textsc{i} map shows diffuse emission, with no clear shell-like morphology or unambiguous spatial correspondence with the radio boundary of the remnant. The $^{12}$CO emission appears more structured, with an enhancement adjacent to the northern edge of the remnant and a morphology that partially traces the radio boundary of the SNR and a portion of the nearby \hii. From this molecular clump, some fainter, elongated $^{12}$CO structures extend across the northern interior of the remnant and intersect the \hii, roughly aligned with the filamentary radio structures observed in the same area.

Despite these indications, the emission peak is not dominant in either the $^{12}$CO or H\textsc{i} profile, but remains relatively weak compared to the overall velocity distribution. In addition, no clear morphological association is observed in either the molecular or H\textsc{i} emission. Therefore, while the gas at this velocity may belong to the same large-scale environment as the SNR, we conclude that the present data do not provide firm evidence for a direct physical interaction.

\subsubsection{\hii\ G343.147$-$00.443} 
\label{subsubsection:HII G343.147-00.443}
We treated the nearby \hii\ as a separate case, since its different kinematic distance \citep{Anderson_2014} indicates no association with the SNR or with the surrounding molecular and H\textsc{i} emission discussed above.
Considering the H\textsc{i} SGPS velocity profile extracted over the \hii\ (Fig.\ref{fig:vel_prfile_HI_HIIreg}, upper panel), we focus on the prominent spectral feature within the velocity range $\sim -39.57$ to $\sim -24.73$ km s$^{-1}$ (shaded area in Fig.\ref{fig:vel_prfile_HI_HIIreg}). We used the updated solar motion parameters from \citet{Reid_2014} (galactocentric radius $R_0 = 8.34$ kpc and circular velocity $V_0 = 241$ km s$^{-1}$) to estimate the kinematic distance of this feature, obtaining $d_{\rm near} = 2.34$–$3.36$ kpc and $d_{\rm far} = 12.60$–$13.62$ kpc. These values are consistent with the near and far distances ($d_{\rm near} = 3.1$ kpc and $d_{\rm far} = 13.2$ kpc) reported for the \hii\ G343.147-00.44 by \citet{Anderson_2014}.

The H\textsc{i} spectrum extracted from the \hii\ shows a relatively symmetric profile, with two peaks of $\sim95$ K and a central minimum of $\sim80$ K at $\sim -32.15$~km s$^{-1}$. Such a morphology may be indicative of emission arising from an expanding structure. 
When an \hii\ is in the expansion
phase \citep{Deharveng_2010}, the H\textsc{i} emission originates in the neutral layer at the interface with the surrounding medium, which expands at typical velocities of a few ~km s$^{-1}$ up to $\sim10$~km s$^{-1}$ (\citealt{Deharveng_2005}, \citealt{Hosokawa_2006}). The contributions from the approaching and receding sides of this H\textsc{i} shell can produce a double-peaked velocity profile separated by $\sim 2v_{\rm exp}$ (typically $\sim10$–$30$ km s$^{-1}$).

To investigate this scenario, we analysed the H\textsc{i} morphology by producing intensity maps integrated over the velocity intervals corresponding to the two peaks at central velocity of $\sim -37.10$ and $\sim -27.21$ km s$^{-1}$, and the central minimum centred at $\sim -32.15$ km s$^{-1}$ (shaded regions in Fig.~\ref{fig:vel_prfile_HI_HIIreg}, upper panel). The resulting maps (Fig.~\ref{fig:vel_prfile_HI_HIIreg}, central panels) reveal a coherent velocity-dependent structure: the blueshifted emission traces H\textsc{i} material located on the foreground of the \hii\ (left panel), while the central velocity range outlines a cavity-like structure roughly coincident with the ionized region (central panel). The redshifted component appears to trace far-side H\textsc{i} gas (right panel).
Overall, the spatial and kinematic properties of the H\textsc{i} emission are consistent with a cavity-like structure possibly associated with an expanding nebula, although projection effects and the irregular morphology prevent a definitive identification of a well-defined, symmetric shell.  

We also investigated the molecular material towards the \hii\ G343.147$-$00.443 through the  $^{12}$CO ($J = 1-0$) maps provided by MOPRA. In Fig.\ref{fig:vel_prfile_HI_HIIreg} (upper panel), we show the $^{12}$CO velocity profile (red line) across the \hii\ in comparison with that of the H\textsc{i}, revealing some difference in their kinematic behaviour. Both profiles revealed structures compatible with the \hii\ distance, suggesting that the atomic and molecular gas may be part of the same large-scale environment. However, while the H\textsc{i} emission shows structured variations within the velocity interval, most of the CO emission is confined to a narrower component centred at $\sim -26$ km s$^{-1}$. This indicates that both tracers are associated with the same physical structure, but probe different phases of the ISM.

The absence of clear expansion signatures in the CO spectrum suggests that the denser molecular gas either remains largely unaffected by the expansion of the \hii\ or has been partially photodissociated. In contrast, the H\textsc{i} emission traces the lower-density neutral gas, which is more efficiently displaced by the ionized region, producing the observed cavity-like morphology. This behaviour is consistent with a scenario in which the \hii\ remains associated with its parent molecular cloud and is likely beginning to modify the surrounding neutral medium, without yet producing a clear expansion signature in the molecular gas. To further explore this scenario, we studied the morphological characteristics of the $^{12}$CO emission by producing three images integrated on the same velocity intervals used for the H\textsc{i} images  (Fig.\ref{fig:vel_prfile_HI_HIIreg}, bottom panels). The CO intensity maps do not reveal a clear cavity structure associated with an expanding molecular shell. Instead, the CO emission exhibits a clumpy distribution, predominantly concentrated towards the northwestern side of the \hii. 
The relatively symmetric H\textsc{i} velocity profile, compared to the asymmetric distribution of the CO emission, further supports a scenario in which the atomic gas traces a more globally redistributed component, while the denser molecular material is unevenly distributed and preferentially concentrated on one side of the \hii, possibly along the foreground portion of the parental cloud.

\section{Discussion}
\label{sec:discussion}
\subsection{Insights on the evolutionary stage of G343.1$-$00.7}
\label{sec:Insights on the evolutionary stage of G343.1$-$00.7}

We used the results from our radio analysis to infer the evolutionary status of G343.1$-$00.7. 
We followed the complementary theoretical approaches proposed by \citet{Urosevic_2020, Urosevic_2022} and \citet{Leahy_2019}. A brief description of the two methods and their application to G343.1$-$00.7 is provided below.

The approach described by \citet{Urosevic_2020, Urosevic_2022} relies on three key diagnostics, all of them based on observational radio parameters:
(i) the position of the SNR in the radio surface brightness–to–diameter ($\Sigma$–D) diagram, relative to the theoretically derived SNR $\Sigma$–D tracks;
(ii) the slope and shape of the radio spectrum; and
(iii) the magnetic field strength, estimated through equipartition (eqp) calculations\footnote{\url{https://poincare.matf.bg.ac.rs/~arbo/eqp/}}.

\begin{figure}
    \centering

    \includegraphics[width=\columnwidth]{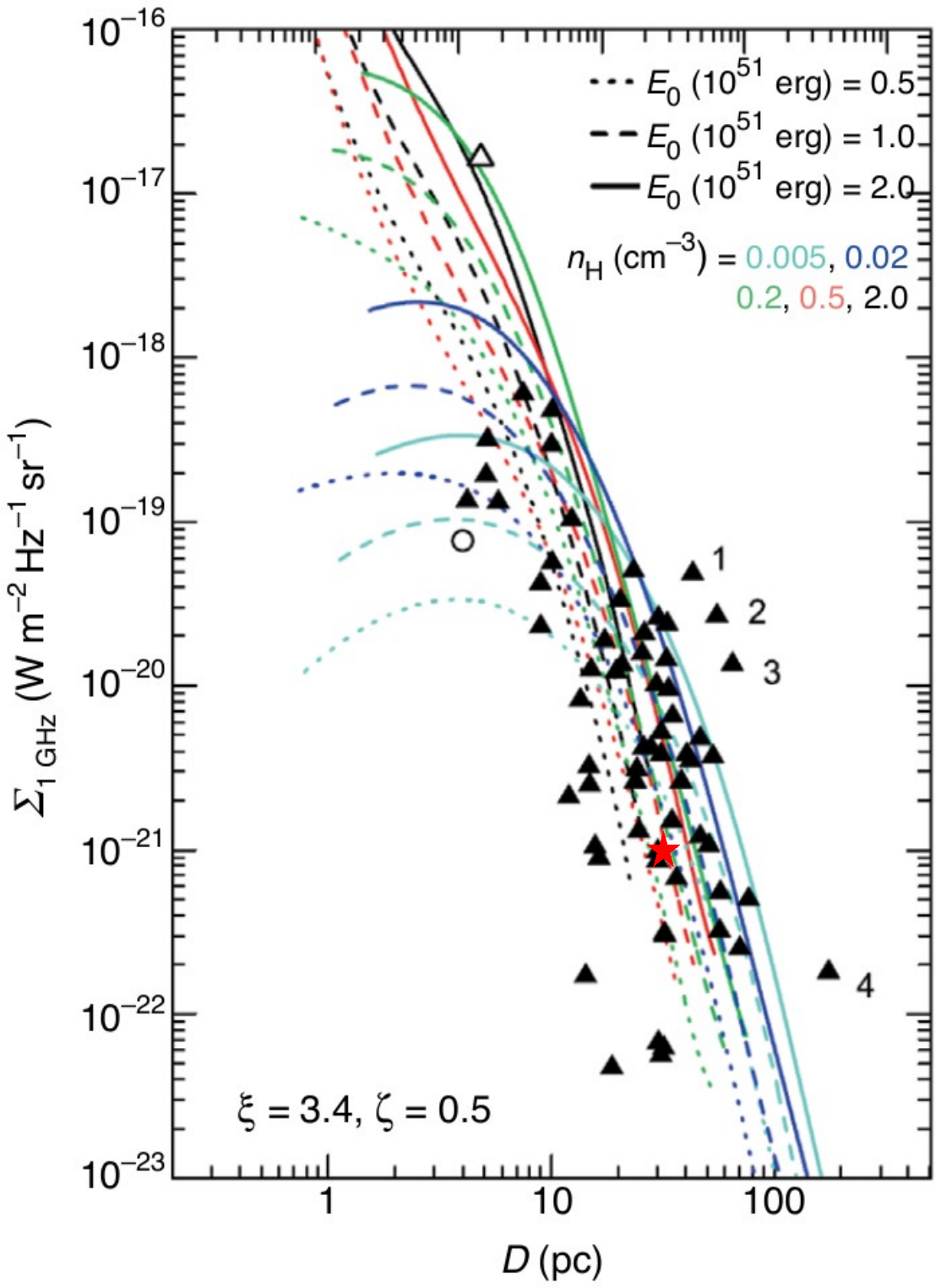}

    \caption{ Radio surface brightness–to–diameter diagram showing the theoretically derived evolutionary paths (lines) and the observed positions of 65 Galactic SNRs (black triangles) from \citet{Pavlovic_2018}, adapted from their Fig.~3. The position of G343.1$-$00.7 is marked with a red star, located in the lower region of the diagram, where the evolutionary tracks reach their steepest slope, corresponding to the full Sedov phase of the SNR evolution.
    }
    \label{fig:Sigma_D}
    
\end{figure}

For the first point, we calculated the radio surface brightness of G343.1$-$00.7 at 1~GHz as $\Sigma \sim 1.5 \times 10^{-21}$~W~m$^{-2}$~Hz$^{-1}$~sr$^{-1}$, using a flux density of 8.16~Jy at 1~GHz and an angular diameter of $\sim$28.6~arcmin. 
Assuming a distance of 4.9~kpc \citep{Ranasinghe_2022}, this yields an intrinsic diameter of $D \sim 41$~pc. These values place G343.1$-$00.7 in the lower portion of the $\Sigma$–D diagram, where evolutionary tracks become steep—characteristic of the full Sedov–Taylor (ST) phase of SNR evolution. In Fig.~\ref{fig:Sigma_D}, we report the $\Sigma$–D diagram from Fig.~1 of \citet{Pavlovic_2018}, which includes theoretically derived evolutionary tracks, overlaid with the positions of 65 Galactic SNRs with known distances. We added G343.1$-$00.7 to this diagram, marking its position with a red star. The position of the SNR on the $\Sigma$–D evolutionary tracks suggests that it is evolving within a relatively dense environment ($0.2$–$0.5$~cm$^{-3}$), assuming a canonical explosion energy of $1\times10^{51}$~erg. 

Concerning the second diagnostic, the integrated radio spectrum of G343.1$-$00.7 is well described by a single power law with a spectral index of $\alpha = -0.5$, showing no indication of curvature at high frequencies. According to Table 1 in \citet{Urosevic_2022}, this spectral behaviour is consistent with either a young SNR evolving under the test-particle DSA scenario or an evolved SNR under the DSA assumptions. Considering its low surface brightness and large physical size, G343.1$-$00.7 is more plausibly associated with the latter class, similar to other evolved remnants such as the Monoceros and Lupus Loop SNRs \citep{Urosevic_2022}.

To explore the third diagnostic, we first estimated the magnetic-field
strength using the electron eqp formalism developed by
\citet{Urosevic_2018} and \citet{Filipovic_2023}. In this approach,
the magnetic-field energy density is assumed to be comparable to the
energy density of CR electrons
($\epsilon_{e}\simeq\epsilon_{B}=B^2/8\pi$).
Using 3D hydrodynamic simulations coupled with a non-linear diffusive
shock acceleration model, \citet{Urosevic_2018} showed that, in evolved
SNRs during the ST phase, the ratio between the CR electron and magnetic-field energy densities does not change significantly during the remnant evolution. In their simulations,
the ratio $\epsilon_e/\epsilon_B$ ranges approximately between 0.1 and
0.9, with an average value close to 0.5. In particular, for the evolved
SNR HB3 they obtained $\epsilon_e/\epsilon_B \sim 0.7$, with only a
moderate variation during its evolution.
As input parameters, we adopted the spectral index of $\alpha \sim -0.5$ derived from our analysis, a flux density of 8.4~Jy at 0.944~GHz, a distance of 4.9~kpc, and an angular radius of 14.3~arcmin. We assumed a filling factor of $f = 0.1$\footnote{We repeated the calculation for values of $f$ between 0.1 and 0.9, obtaining an estimated minimum-energy magnetic field between  23.8~$\mu$~G and 12.8~$\mu$~G. Although the assumed filling factor introduces a systematic decrease in the inferred field strength, the magnetic field remains of the order of 10~$\mu$~G.
Such values remain consistent with a compression of the ambient magnetic field at the strong non-radiative shocks (by a minimal factor of 2-3), while ruling out magnetic-field amplification, as expected for evolved SNRs.
Therefore, the main conclusions regarding magnetic-field amplification are not affected by the adopted filling factor.}, a compression ratio $\sigma \simeq 4$, and injection parameter $\xi \simeq 4$, consistent with relatively slow non-radiative shocks. A shock velocity of $500$~km\,s$^{-1}$ was adopted, representative of slow non-radiative shocks. We adopted an electron-to-proton temperature ratio of $T_{\mathrm{e}}/T_{\mathrm{p}} = 0.8$ \citep{Ghavamian_2007}.
The electron eqp calculation yields a magnetic-field strength
of $B\sim25~\mu$G.

The adopted velocity of $500$~km\,s$^{-1}$ certainly does not favour eqp between CR protons and the magnetic field, and even eqp between CR electrons and the magnetic field may not be fully justified.  
Since G343.1$-$00.7 is inferred to be an evolved remnant, we adopted the more general constant-partition approximation
discussed by \citet{Arbutina_2012}  and \citet{Urosevic_2018}. 
By introducing a constant partition with $\epsilon_e/\epsilon_B \sim 0.5$ into our eqp calculation, we obtained a revised magnetic-field estimate of the order of $30~\mu$G.
Given the approximate, order-of-magnitude nature of this method, the
difference between the pure eqp and constant-partition estimates is not significant and does not affect our physical interpretation.

Our magnetic-field strength estimation is consistent with the expected compression of the interstellar magnetic field by a minimal factor of 2--3 (at the strong non-radiative shocks; in the radiative shocks, compression is significantly higher), assuming $B_0\sim7-9~\mu$G, comparable to an average Galactic ISM magnetic field of $\sim5~\mu$G \citep{Wielebinski_2005}. This value is also consistent with a leptonic-dominated emission scenario \citep{Loru_2018}. The absence of detectable $\gamma$-ray emission from G343.1$-$00.7 further supports this interpretation.
Under the electron eqp assumption, the total electron energy is estimated to be \(E_{\mathrm{e}} = 2.4\times10^{48}\)~erg  and proton energy \(E_{\mathrm{p}} = 5.8\times10^{49}\)~erg. This value is consistent with the standard scenario in which up to \(\sim10\%\) of the supernova kinetic energy (\(E_{\mathrm{SN}} \sim 10^{51}\)~erg) is transferred to CR particles and only a few percent of the CR energy is carried by electrons \citep{SNR_book_2020}.
When compared with the $B$ and $E_{\mathrm{e}}$ values reported by \citet{Urosevic_2018} for a sample of 65 well-studied SNRs, G343.1$-$00.7 falls within the typical range expected for evolved remnants. In contrast, younger SNRs such as Cas~A \citep{Urosevic_2018} and G1.9+0.3 \citep{Pavlovic_2017} exhibit significantly higher magnetic-field strengths, often reaching several hundred micro Gauss.
By combining the results from the three independent radio diagnostics, the model suggests that G343.1$-$00.7 is likely an evolved SNR currently in the late ST phase evolving in a relatively dense environment.

We further investigated the evolutionary phase of G343.1$-$00.7 by applying the models presented by \citet{2019AJ....158..149L}. These models provide the evolution of the SNR radius, shock temperature, X-ray emission measure, and other physical quantities as a function of time for a given set of input parameters. The latter include the explosion energy $E_0$, ISM density $n_0$, and ejecta mass $M_{\rm ej}$.
The mean explosion energy of Galactic SNRs was shown to be $3\times10^{50}$ erg by \cite{2020ApJS..248...16L} with a 1 $\sigma$ dispersion of a factor of 3. 
The ambient density at the position of the SNR was estimated using the Galactic density model of \cite{2006A&A...459..113M}.
For G343.1$-$00.7, we used a distance of 4.9 kpc, giving the SNR a Galacto-centric distance of 3.9 kpc and height above the plane of 60 pc. 
The result is a density of $H_2$ of 0.424 cm$^{-3}$ and of H\textsc{i} of 0.181 cm$^{-3}$, for a total density of $2 n_{H_2}+ n_{HI}$= 1.03 cm$^{-3}$.
For a late stage of evolution, the amount of ejected mass does not have a strong effect on the evolution; thus, we take it to be 1.4 M$_{\odot}$.

For the above input parameters $E_0$=$3\times10^{50}$ erg, ISM density $n_0$=1.0 cm$^{-3}$, and ejected mass $M_{ej}$=1.4 M$_{\odot}$, we calculated the SNR evolution to reach its current radius of 20.5 pc.
The resulting age is 88500 yr, so it is beyond the end of the ST stage and into the pressure-driven shell (PDS) phase, with a ST-to-PDS transition time of 10,300 yr.
The X-ray emission measure is $\sim3\times10^{58}$ cm$^{-3}$ after correcting for cooling in the PDS phase with a low shock temperature of $1.4\times10^5$ K.
If we use $E_0$=$1\times10^{51}$ erg, 
the resulting age is 38500 yr, placing it in the PDS phase, with a transition time from ST to PDS of 13,300 yr. 
The X-ray emission measure is $\sim1\times10^{60}$ cm$^{-3}$ after correcting for cooling in the PDS phase with a shock temperature of $8.3\times10^5$ K.

The fact that no X-ray emission was detected from G343.1$-$00.7 by eROSITA is consistent with the SNR being in a late phase of evolution.
The eROSITA upper limit to the X-ray flux is $2\times10^{-14}$ erg cm$^{-2}$ s$^{-1}$ in the low energy band (0.2 to 2.3 keV) and is $4\times10^{-13}$ erg cm$^{-2}$ s$^{-1}$ in the high energy band (2.3 to 5.0 keV).
The estimated column density towards G343.1$-$00.7, based on the model of \cite{2006A&A...459..113M} is $2\times10^{21}$ cm$^{-2}$, which can absorb soft X-rays. 
The above model with $E_0$=$3\times10^{50}$ erg yields a low enough emission measure and shock temperature to be consistent with the non-detection, while the higher-energy model with $E_0$=$1\times10^{51}$ erg is not consistent with the X-ray non-detection.

\subsection{Discerning emission processes across the remnant}
\label{sec:Discerning emission processes across the remnant}
Our spectral-index map (Fig.\ref{fig:spix}) reveals significant spectral variations indicative of different shock conditions and emission processes across the shell structure of G343.1$-$00.7. While the north-northwestern shell and the southern edge show mean spectral indices of $\sim -0.62$ and $\sim -0.39$, respectively, consistent with standard synchrotron emission from shock-accelerated electrons via DSA, the southeastern region exhibits significantly flatter spectral indices of $\sim -0.27$.
From a morphological point of view, this region forms part of the remnant shell and is characterized by thin filamentary radio structures, at least three of which are particularly bright and well resolved in our ASKAP-EMU image (Fig.~\ref{fig:brightness_maps}). 
Such thin, bright radio filaments are generally interpreted as the result of shock interaction with an inhomogeneous medium, where density enhancements lead to strong compression of the post-shock plasma and amplification of the magnetic field, leading to localized enhancements of the synchrotron emissivity. 

If the SNR blast wave expands in a high-density environment with sufficient velocity to dissociate and ionize the gas, the shock may transition towards radiative conditions, potentially producing significant thermal bremsstrahlung emission at radio continuum frequencies (\citealt{Onic_2012}; \citealt{Urosevic_2005}).
When these conditions occur locally in an inhomogeneous medium containing dense cloudlets or strong density gradients due to the proximity of molecular clouds, the affected regions may evolve more rapidly, reaching the radiative phase while the remnant as a whole remains in the adiabatic phase \citep{Reynolds_2008}.
In these cases, local spectral flattening may arise from a contribution of thermal bremsstrahlung emission at radio continuum frequencies, which is characterized by a power-law spectrum with spectral index $\sim -0.1$ \citep{Urosevic_2005}. Additional signatures indicating a competing contribution between thermal bremsstrahlung and steeper synchrotron emission include \citep{Onic_2012}: spectral flattening at high frequencies (concave-up curvature above $\sim1$~GHz); intrinsic thermal absorption at low radio frequencies; and morphological association with high-density regions lacking strong linear polarization.

In Sect.~\ref{HI environment}, we investigated the environmental conditions around G343.1$-$00.7, revealing the presence of both H\textsc{i} and $^{12}$CO emission in positional and kinematic correspondence with the remnant. 
By considering the H\textsc{i} emission integrated over the velocity interval from $-80.0$ to $-68.60$~km\,s$^{-1}$ (see Sect.\ref{subsubsection:SNR G343.1-00.7}) and under the assumption of optically thin H\textsc{i} emission, we estimated an H\textsc{i} column density towards the southeastern shell region of $N_{\mathrm{HI}} \sim 3 \times 10^{20}$~cm$^{-2}$.
To estimate the ambient atomic density, we assumed an approximately spherical geometry for G343.1$-$00.7 and adopted a distance of 4.9~kpc. We further considered a characteristic line-of-sight depth comparable to the projected radius of the remnant, corresponding to \(L \sim 20\)~pc (\(\sim 6.17\times10^{19}\)~cm). Under these assumptions, we derive an order-of-magnitude estimate of \(n_{\mathrm{HI}} \sim 4.8\)~cm$^{-3}$. This value is consistent with the typical density range expected for a relatively dense warm ISM (\(n \sim 1\)--\(10\)~cm$^{-3}$), satisfying the necessary condition for the production of significant thermal radio emission through the bremsstrahlung process in evolved SNRs \citep{Urosevic_2005}.

\begin{figure*}
    \centering
    \includegraphics[width=18cm]{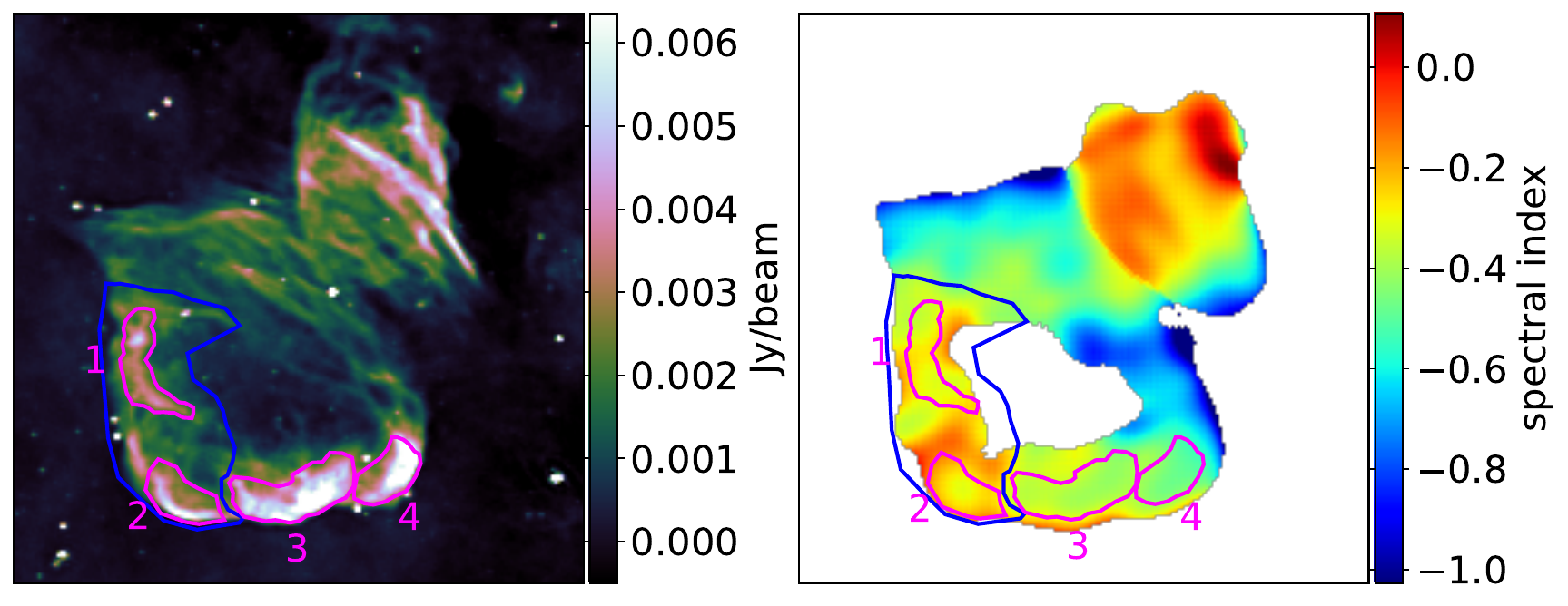}
    \includegraphics[width=4.5cm]{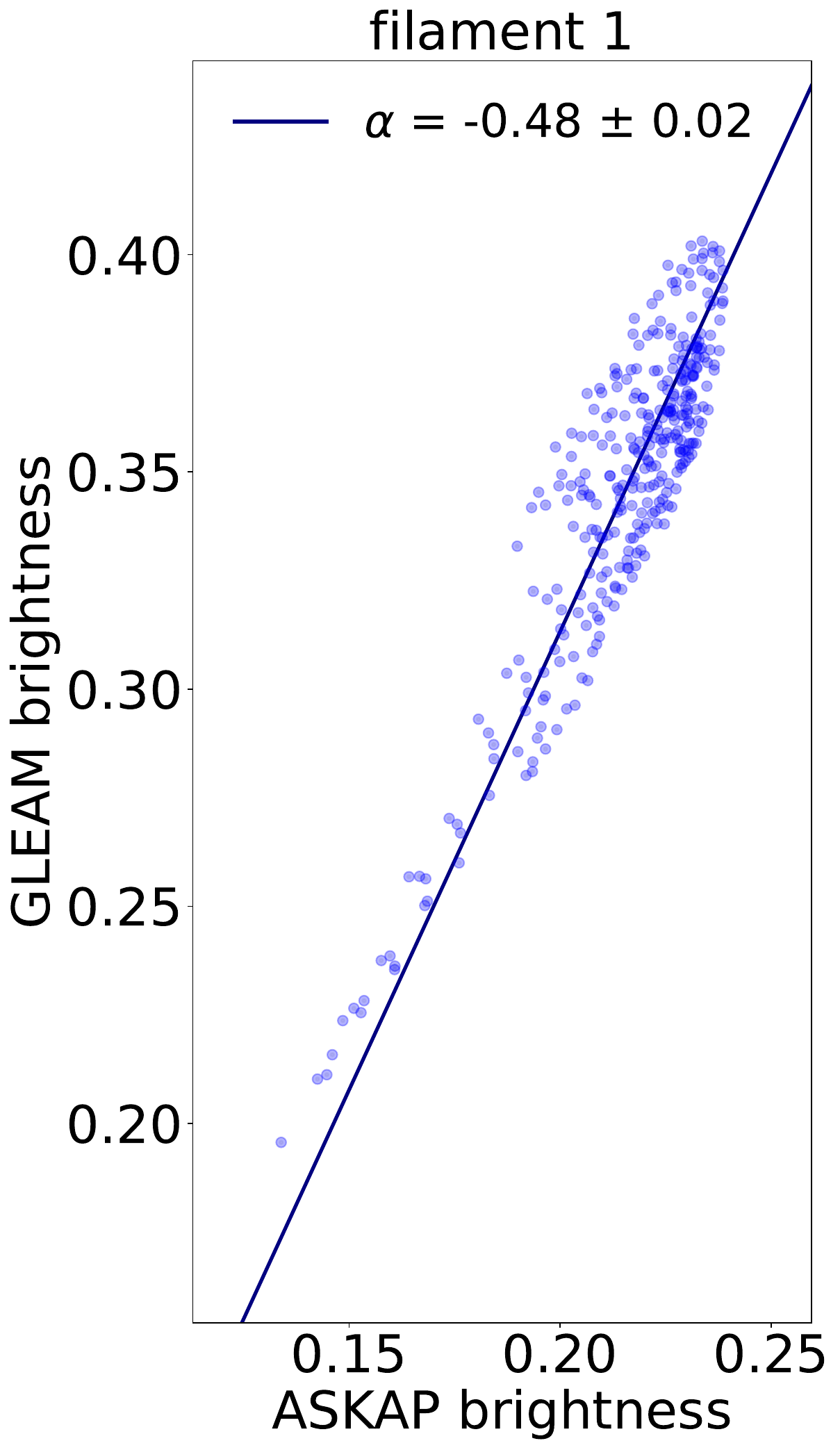}
    \includegraphics[width=4.5cm]{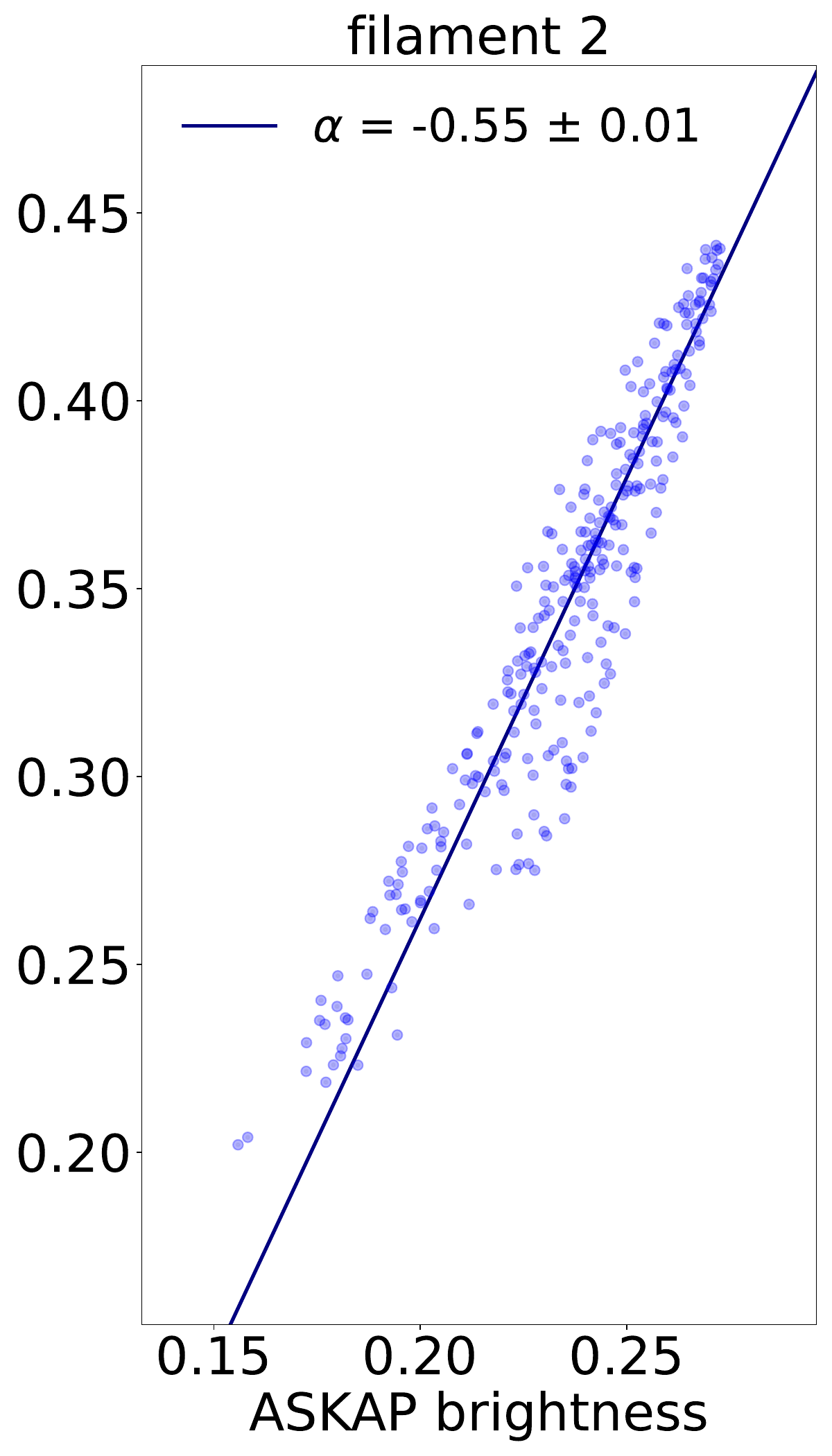}
    \includegraphics[width=4.5cm]{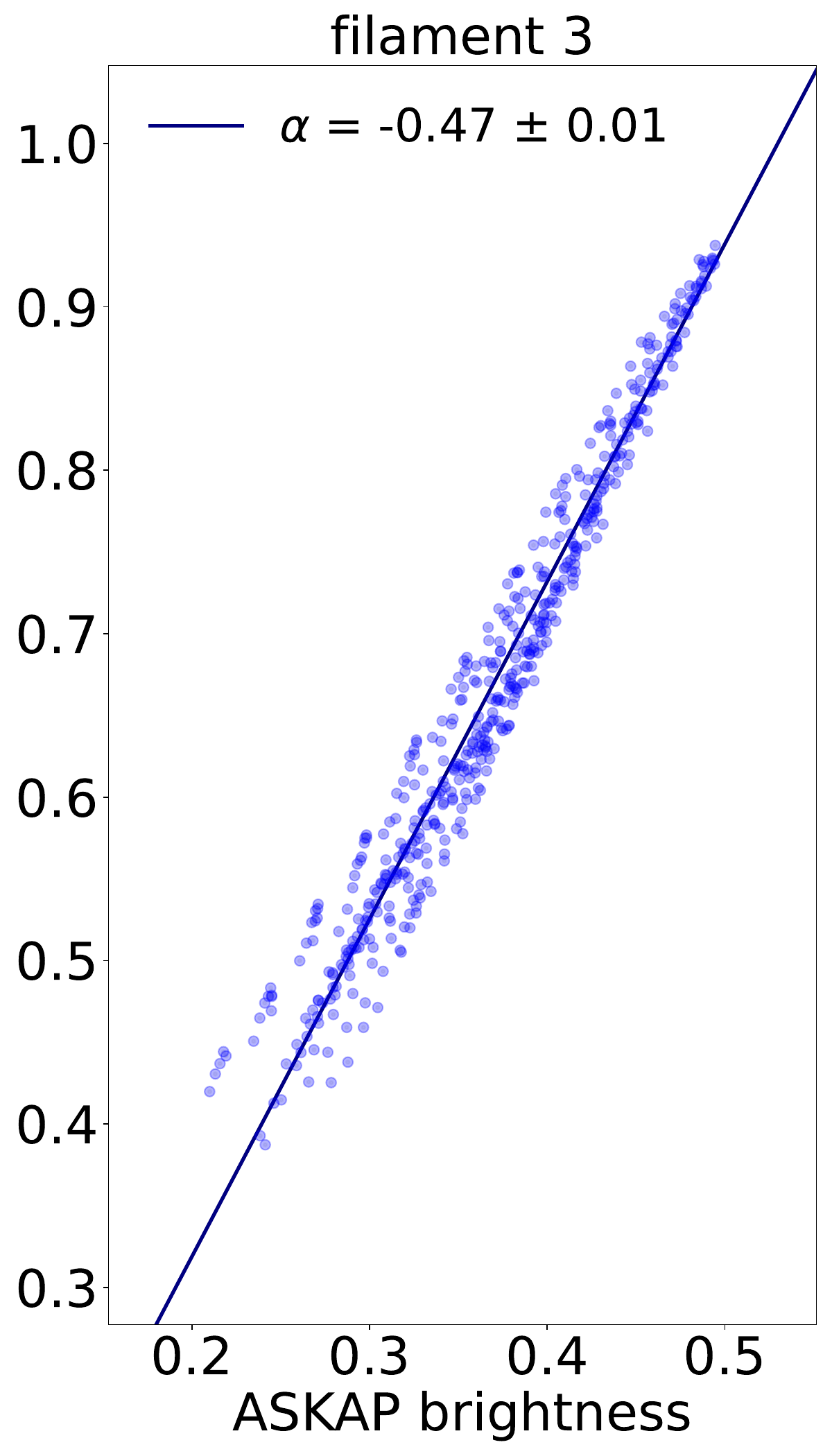}
    \includegraphics[width=4.5cm]{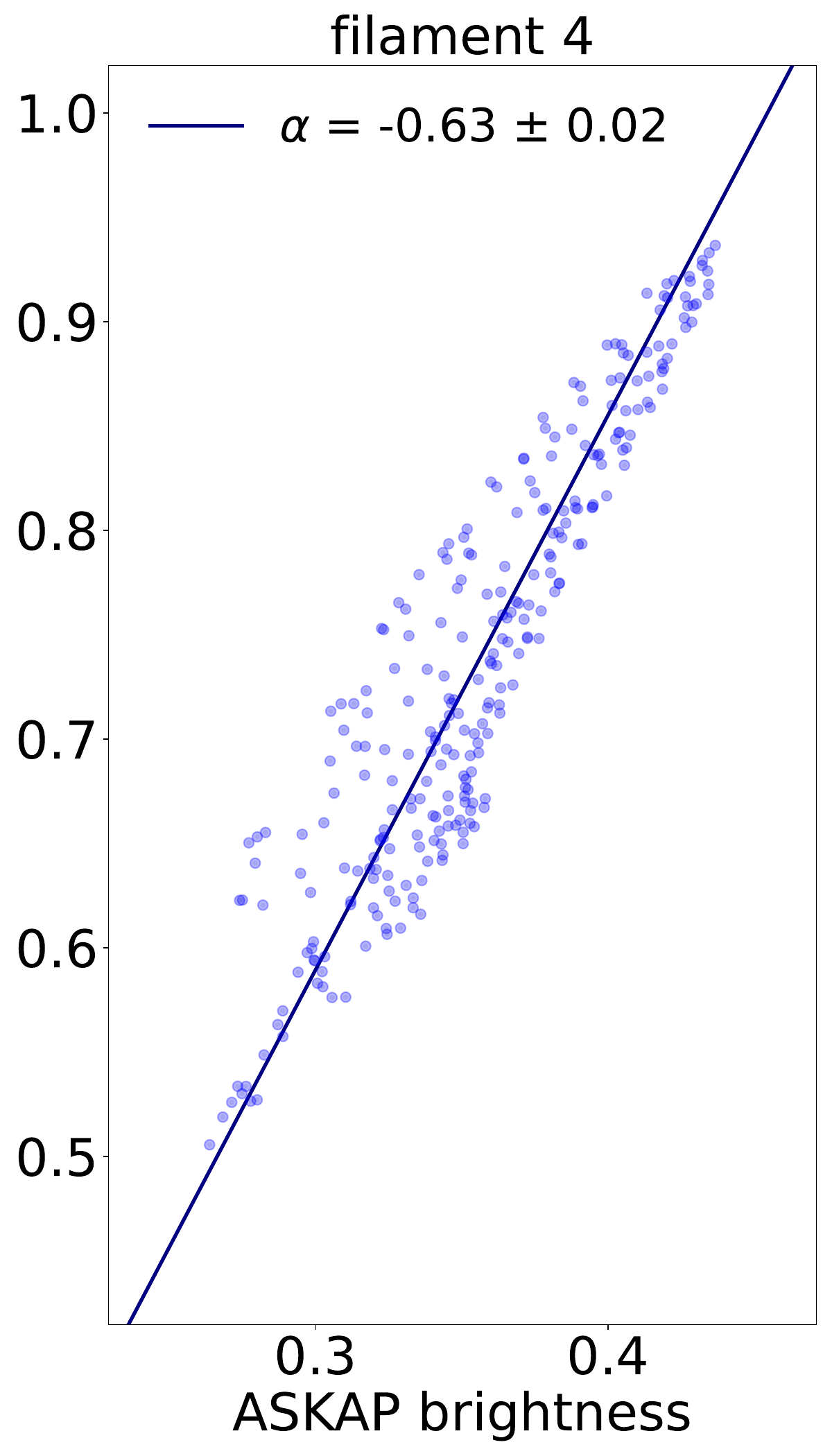}
    
    \caption{Spatially resolved spectral index study of the southeastern region of SNR G343.1$-$00.7. 
    \textit{Top:} EMU image of G343.1$-$00.7 at 0.944~GHz and the corresponding GLEAM-EMU 0.200--0.944~GHz spectral-index map. The blue region encloses the southeastern shell, while the magenta regions indicate the filamentary structures analysed in this work. 
    \textit{Bottom:} BB-plots between 0.200 and 0.944~GHz for the four selected filaments. Brightness values are expressed in Jansky per beam.}
    \label{fig:radiative_filaments}
\end{figure*}

Starting from the possibility of coexisting synchrotron and thermal bremsstrahlung emission in this region, we performed a brightness--brightness analysis of the two brightest filaments in the southeastern shell using the EMU image at 0.944~GHz and the GLEAM image at 0.200~GHz. 
Since this method is largely insensitive to zero-level offsets in the input images, it provides a more reliable estimate of the intrinsic spectral index of compact bright structures while minimizing the contribution from the surrounding diffuse emission. Using background-subtracted radio images allows us to distinguish the spectral properties of the bright filaments from those of the co-spatial diffuse component.
The two filaments are outlined by the polygonal regions labelled 1 and 2 in Fig.~\ref{fig:radiative_filaments} and have mean spectral index values of $\sim -0.30$ and $\sim -0.24$, respectively, derived from the spectral-index map. For comparison, the same analysis was performed on two bright filaments located in the southwestern shell (regions 3 and 4 in Fig.~\ref{fig:radiative_filaments}), whose spectral-index map yields steeper mean spectral indices of $\sim-0.38$ and $\sim-0.46$, respectively. 
The corresponding BB-plots are shown in the bottom panels of Fig.~\ref{fig:radiative_filaments}. 

In all cases, the data points display relatively narrow linear distributions, yielding spectral indices significantly steeper than those inferred from the spectral-index map. This behaviour suggests that the compact filamentary structures are dominated by synchrotron emission, which is more reliably highlighted by the BB analysis, while the flatter spectral component is likely associated with diffuse large-scale emission that is effectively treated as a background contribution in the BB-plots.
This result is consistent with a scenario in which the spectral flattening observed in the southeastern shell is produced by an additional thermal bremsstrahlung component arising from diffuse post-shock plasma, whereas the bright filamentary structures are predominantly tracing synchrotron emission associated with the shock front.

A similar behaviour is observed for the southwestern filaments, for which the spectral indices derived from the BB-plots are also steeper than the mean values inferred from the spectral-index map. This suggests that the bright filamentary emission in both regions is primarily dominated by synchrotron radiation, whereas the flatter spectral-index-map values are likely influenced by diffuse large-scale emission from the surroundings. However, unlike the southeastern region, the southwestern filaments do not exhibit the same degree of spectral flattening nor comparable environmental indications of a possible thermal contribution. In this case, the moderate flattening suggested by the spectral-index map is more plausibly associated with contamination from diffuse synchrotron emission, making the evidence for thermal bremsstrahlung emission or radiative shock conditions less compelling than in the flatter southeastern filaments, although such a contribution cannot be entirely excluded.

To further investigate the possible presence of thermal emission associated with radiative or partially radiative shocks, we searched for signatures of thermal bremsstrahlung in the integrated spectrum of the southeastern region of the remnant. We calculated the integrated flux densities within a region enclosing the filamentary structures observed in the southeastern shell using the GLEAM and EMU maps (blue polygonal region in Fig.\ref{fig:radiative_filaments}, top panels). 
We fitted the resulting measurements using a model including both non-thermal synchrotron ($S_{\mathrm{NT}}$) and thermal bremsstrahlung ($S_{\mathrm{T}}$) emission, where the integrated spectrum is described as the sum of the two components:
\begin{equation}
    S_{\nu}= S_{\mathrm{NT}}\nu^{\alpha} + S_{\mathrm{T}}\nu^{\alpha_{\mathrm{T}}}
    \quad (\mathrm{Jy}).
\end{equation}

We fixed the non-thermal spectral index to $\alpha_{\mathrm{NT}}=-0.5$, corresponding to the value derived for the whole SNR (Sect.~\ref{sec:Results}) and also consistent with the spectral indices obtained for the two bright filaments from the BB-plot fits. The thermal spectral index was fixed to $\alpha_{\mathrm{T}}=-0.1$, as expected for optically thin thermal bremsstrahlung emission.

\begin{figure}
    \centering
    \includegraphics[width=\columnwidth]{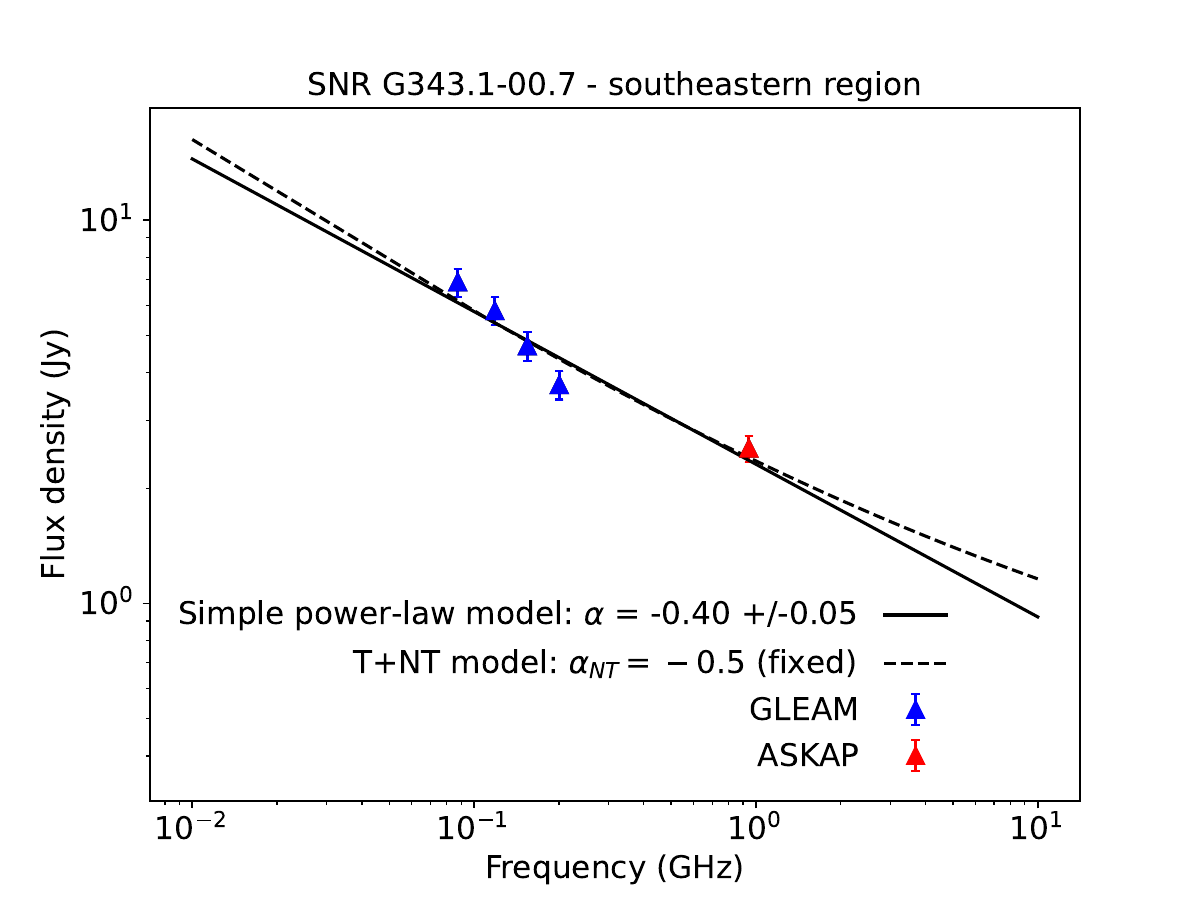}
    \caption{Integrated spectrum of the southeastern shell region of G343.1$-$00.7. The dashed line shows the fit derived from the combined non-thermal plus thermal model, while the solid line indicates the purely non-thermal fit. }
    \label{fig:SE_SED}
\end{figure}

Figure~\ref{fig:SE_SED} shows the integrated spectrum of the southeastern shell region together with the fit obtained using the combined non-thermal plus thermal model (solid line), compared with the fit derived from a purely non-thermal model ($S_{\nu}\propto \nu^{\alpha}$). The fit parameters and corresponding results are reported in Table~\ref{tab:sed_fit_results}.
From the combined model fit, we inferred a thermal contribution at 1~GHz of $\sim37\%$, ranging between about $23\%$ and $50\%$. This estimate is consistent with the thermal fractions reported by \citet{Onic_2012} for candidate radio thermally active SNRs.

The combined thermal and non-thermal model yields a slightly lower Bayesian information criterion (BIC; \citealt{Schwarz_1978}) than the purely non-thermal model. However, this difference ($\Delta\mathrm{BIC} = \mathrm{BIC}_{\rm NT} - \mathrm{BIC}_{\rm NT+TH} \simeq 1$) is too small to provide substantial evidence in favour of the combined model or to establish the presence of a significant thermal component in G343.1$-$00.7. Additional observations, particularly in the 10--30\,GHz frequency range, are required to better constrain the spectral decomposition and assess the significance of any thermal emission.
We point out that sensitive, high-resolution radio observations over a broad frequency range, from low ($\lesssim 100$ MHz) to high frequencies ($\gtrsim 10$ GHz), combined with polarization information and higher-resolution spatially resolved spectral analysis, are crucial to disentangle the spectral behaviour across the southeastern region of G343.1$-$00.7 and constrain the relative contributions of synchrotron and possible thermal emission components.

\subsection{Overall interpretation and future perspectives}
\label{subsec: Overall interpretation and future perspectives}

As a final result, the two models discussed in Sect.\ref{sec:Insights on the evolutionary stage of G343.1$-$00.7} place G343.1$-$00.7 between the late ST phase and the early PDS phase. We do not consider the differences between the two results to be contradictory, as they arise from different and complementary methods, but rather as an indication that different regions of the remnant may be at different evolutionary stages. This interpretation is further supported by our findings on the radiative nature of the bright filaments along the southeastern edge of the shell. Such local differences in the evolutionary state are expected for SNRs evolving in an inhomogeneous ISM, where portions of the blast wave may become radiative earlier in regions of enhanced ambient density, as observed, for example, in the SNR Cygnus Loop \citep{Urosevic_2026}.

Overall, this work provides the first comprehensive radio picture of G343.1$-$00.7, establishing a solid observational framework for future multiwavelength studies. In particular, high-resolution radio observations at higher frequencies are crucial for firmly constraining the relative contribution of the different emission processes across the remnant shell. 
In this context, the three-band ASKAP (0.9, 1.25, and 1.55~GHz) image shown in Fig.~\ref{fig:ASKAP TRI-BAND}, although suitable only for qualitative analysis because of the limitations discussed in Sect.~\ref{sec:Radio data}, clearly illustrates the potential of multi-band, high-resolution radio imaging for spatially resolved spectral studies. The RGB image immediately distinguishes the non-thermal emission associated with G343.1$-$00.7 (displayed in red and white) from the thermal emission of the nearby \hii\ (displayed in blue). Moreover, the bright filamentary structures attributed to synchrotron emission remain morphologically consistent across the three bands, making it possible to distinguish the remnant from the overlapping \hii\ even in the northwestern part of the complex.

High-resolution, multi-band radio images obtained simultaneously with the same instrument and then matched in angular resolution and pixel scale remain extremely rare. Our qualitative analysis demonstrates their strong potential for spatially resolved spectral studies of Galactic sources, particularly in complex environments where SNRs overlap unrelated thermal structures or exhibit localized variations in their emission properties, as in the case of G343.1$-$00.7. Future ASKAP observations covering all three frequency bands will enable the methodology presented in this work to be refined and applied to a much larger sample of Galactic SNRs.
\begin{table*}[ht]
\centering
\caption{Spectral fitting parameters for the southeastern shell region.}
\label{tab:sed_fit_results}
\begin{tabular}{lcccccc}
\hline
Model &
$\alpha$ &
$S_{\mathrm{1\,GHz}}^{\mathrm{NT}}$  (Jy)&
$S_{\mathrm{1\,GHz}}^{\mathrm{T}}$  (Jy)&
$\frac{S^T}{S^{NT}+S^T}$ &
$\chi^{2}$/dof & BIC  \\
\hline

Purely non-thermal
& $-0.40 \pm 0.05$
& \dots
& \dots
& \dots
& 2.66 & 11.20 \\

Thermal + non-thermal
& $-0.50$ (fixed)
& $1.49 \pm 0.22$
& $0.87 \pm 0.39$
& $0.37 \pm 0.14$
& 2.31 & 10.16 \\

\hline
\end{tabular}
\tablefoot{$\frac{S^T}{S^{NT}+S^T}$ represents the fractional thermal contribution to the total emission.}
\end{table*}

\section{Summary}
\label{sec:summary}
We presented new high-resolution radio-continuum observations of the SNR G343.1$-$00.7 obtained with ASKAP-EMU at $0.944$~GHz, providing an unprecedentedly detailed view of the remnant morphology.
\begin{enumerate}
    \item By combining our flux-density measurements from ASKAP-EMU and 0.088--0.200~GHz MWA-GLEAM data with those available in the literature, we derived an integrated spectral index of $\alpha=-0.50\pm0.01$, with no evidence of a low-frequency turnover.

    \item The combination of the 0.944~GHz EMU and 0.200~GHz GLEAM images allowed us to produce the first spectral-index map of G343.1$-$00.7.
    Together with infrared observations and brightness--brightness analysis, these data enabled us to disentangle the synchrotron-emitting features of the SNR from the thermal emission of the overlapping northern \hii.
    \item We investigated the surrounding ISM using public H\textsc{i} SGPS and $^{12}$CO MOPRA data. The SNR shows only a kinematic association with faint atomic and molecular gas, without clear morphological evidence of interaction. Conversely, the \hii\ shows both morphological and kinematic agreement with the surrounding material, consistent with an early expansion within its parent molecular cloud. This further supports the conclusion that the \hii\ and G343.1$-$00.7 are located at different distances and are physically unrelated.
    \item The spatially resolved spectral analysis revealed significant spectral differences among the brightest radio filaments. In particular, the flatter spectra measured along the southeastern shell, compared with those of the southwestern filaments, are consistent with a possible thermal bremsstrahlung contribution from diffuse post-shock plasma, suggesting that these regions may be entering the radiative phase while the bulk of the remnant remains at an earlier evolutionary stage.
    \item Finally, we used the detailed radio characterization of G343.1$-$00.7 to constrain its evolutionary state. The derived radio properties are consistent with an evolved SNR, with an estimated magnetic field strength of $\sim25\,\mu$G and a total relativistic electron energy of $2.4\times10^{48}$ erg. The two applied models (\citealt{Urosevic_2020}; \citealt{Leahy_2019}) place G343.1$-$00.7 between the late Sedov phase and the early PDS phase, suggesting that different regions of the remnant may be at different evolutionary stages.
     
\end{enumerate}

\begin{acknowledgements}
This scientific work uses data obtained from Inyarrimanha Ilgari Bundara, the CSIRO Murchison Radio-astronomy Observatory. We acknowledge the Wajarri Yamaji People as the Traditional Owners and native title holders of the Observatory site. CSIRO’s ASKAP radio telescope is part of the Australia Telescope National Facility (https://ror.org/05qajvd42). Operation of ASKAP is funded by the Australian Government with support from the National Collaborative Research Infrastructure Strategy. ASKAP uses the resources of the Pawsey Supercomputing Research Centre. Establishment of ASKAP, Inyarrimanha Ilgari Bundara, the CSIRO Murchison Radio-astronomy Observatory and the Pawsey Supercomputing Research Centre are initiatives of the Australian Government, with support from the Government of Western Australia and the Science and Industry Endowment Fund.\\

 DU and BA acknowledge the funding provided by the Ministry of Science, Technological Development, and Innovation of the Republic of Serbia through the contract \# 451-03-33/2026-03/200104. They are also supported through the joint project of the Serbian Academy of Sciences and Arts and Bulgarian Academy of
Sciences—“Detection and Kinematic Characterization of Optical Counterparts to Radio Supernova Remnants.” During the work on this paper BA was supported by the Science Fund of the Republic of Serbia through project \#7337 `Modeling Binary Systems That End in Stellar Mergers and Give Rise to Gravitational Waves' (MOBY). 

CB acknowledges financial support from grant CEX2021-001131-S funded by
MICIU/AEI/10.13039/501100011033 and from grant INFRA24023 (CSIC4SKA) funded by CSIC.
\end{acknowledgements}

\bibliographystyle{aa} 
\bibliography{A_A}

\appendix
\onecolumn
\section{Three-band ASKAP early-science image}
\label{Appendix}

\begin{figure}[!th]
\centering
\begin{minipage}{16cm}
\centering
\includegraphics[width=1.\textwidth]{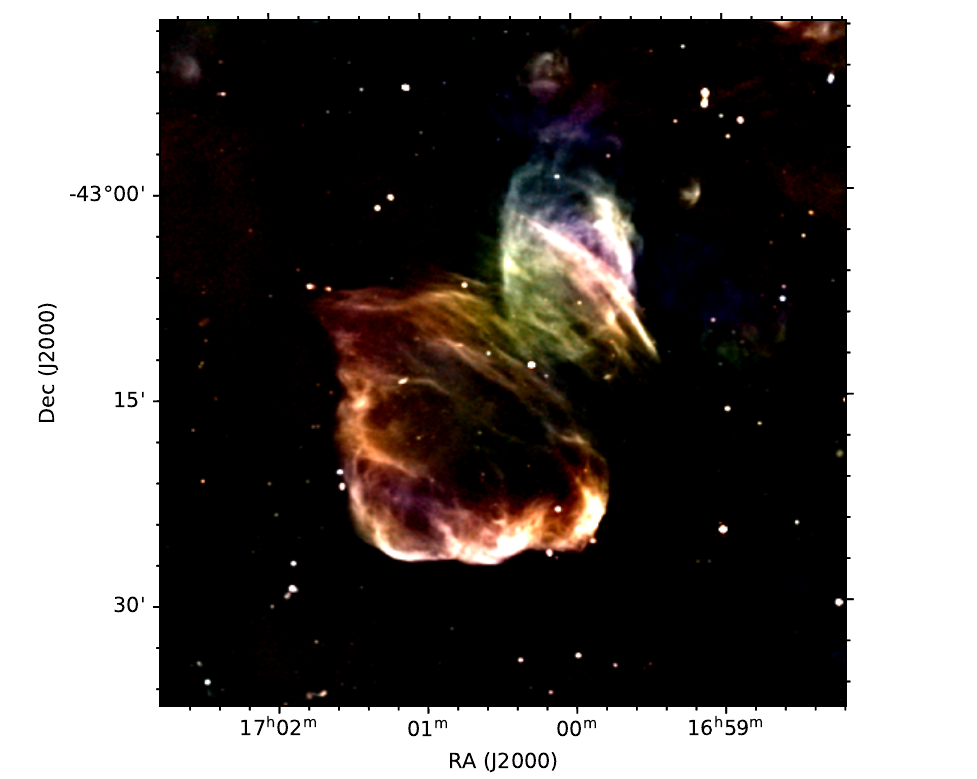}

%captionsetup{width=16cm, justification=justified}
    \caption{
    Three-colour image of the G343.1$-$00.7 complex, including the nearby \hii\ G343.147$-$00.443, obtained from ASKAP Early Science observations. Red: 0.944 GHz (Band 1). Green: 1.25 GHz (Band 2). Blue: 1.55 GHz (Band 3).
    }
    \label{fig:ASKAP TRI-BAND}
    \end{minipage}
\end{figure}

We present a three-colour image of the SNR G343.1$-$00.7 obtained from the ASKAP Early Science observations of the SCORPIO field at central frequencies of 0.9 GHz (Band 1), 1.2 GHz (Band 2), and 1.55 GHz (Band 3). The images at the three frequencies were processed simultaneously and therefore share the same angular resolution and pixel scale. These characteristics allowed us to produce a high-resolution RGB image (pixel size 1.5 arcsec; red: 0.9 GHz, green: 1.2 GHz, and blue: 1.55 GHz; Fig.~\ref{fig:ASKAP TRI-BAND}) that provides a qualitative visualization of the spectral behaviour across the G343.1$-$00.7 complex.

In this image, the non-thermal emission from the SNR appears predominantly red, clearly distinguished from the spectrally flatter thermal emission of the nearby \hii, which appears green-blue and is located along the northwestern edge of the main SNR shell. Moreover, the filamentary structures extending from north-east to south-west exhibit a remarkably uniform morphology and colour across the entire complex, confirming their non-thermal origin and supporting their association with the SNR rather than with the overlapping \hii, even where the two objects are projected onto each other.

\end{document}